\documentclass[showpacs,pre,twocolumn]{revtex4}

\usepackage{graphicx}
\usepackage{amsmath}
\pdfoutput=1

\begin{document}

\title{Density and concentration field description of nonperiodic structures}

\date{\today}

\author{Andreas M.~Menzel}
\altaffiliation[Current address: ]{Max Planck Institute for Polymer Research, P.O.~Box 3148, 55021 Mainz, Germany}
\email[email: ]{menzel@mpip-mainz.mpg.de}
\affiliation{Department of Physics, University of Illinois at Urbana-Champaign, Loomis Laboratory of Physics, 1110 West Green Street, Urbana, IL 61801, USA}

\begin{abstract}
We propose a simple nonlocal energy functional that is suitable for the continuum field characterization of nonperiodic and localized textures. The phenomenological functional is based on the pairwise direction-dependent interaction of field gradients that are separated by a fixed distance. In an appendix, we describe the numerical minimization of our functional. On that basis, we investigate the kinetic evolution of thread-like stripe patterns that are created by the functional when we start from an initially disordered state. At later stages, we find a coarse-graining that shows the same scaling behavior as was obtained for the Cahn-Hilliard equation. In fact, the Cahn-Hilliard model is contained in our characterization as a limiting case. A slight modification of our model omits this coarse-graining and leads to nonperiodic stripe phases. For the latter case, we investigate the temporal evolution of the defects (end points) of the thread-like stripes. In view of possible applications of this functional, we discuss a possible characterization of polymeric systems and vesicles. The statistics of the growth of the thread-like structures is compared to the case of step-growth polymerization reactions. Furthermore, we demonstrate that the functional may be applied for the study of vesicles in a continuum field description. Basic features, such as the tendency of tank-treading in simple shear flows and parachute folding in pipe flows, are reproduced. 
\end{abstract}


\maketitle

\section{Introduction}

For several decades, concentration and density field descriptions have been used to model the behavior of periodic micro- and mesoscopic systems. A prominent example is given by the Ohta-Kawasaki characterization of mesophases as they are observed in block copolymers \cite{ohta1986equilibrium}. This description was derived by averaging over the local positions of the monomers that form the different blocks of the polymers. The procedure led to an effective continuum characterization of the observed periodic structures by the famous Ohta-Kawasaki energy functional. In the case of incompressible diblock copolymers, the order parameter is given by the concentration of one of the two monomer types that form the different blocks. This is why we call this sort of characterization a concentration field description in the following. 

Since then, a lot of studies have demonstrated the effectiveness of the approach. For instance, the kinetic evolution of the periodic mesophase textures from the disordered (spatially homogeneous) state has been investigated \cite{bahiana1990cell}. Also the kinetic evolution from one texture to the other after temperature quenches has been analyzed \cite{nonomura2001growth,yamada2007interface}. Elastic moduli characterizing deformations of the textures were determined \cite{kawasaki1986phase,yamada2006elastic}, and rheological properties of such structures were studied \cite{ohta1993anomalous,doi1993anomalous,drolet1999lamellae, chen2002lamellar,tamate2008structural}. 

In a recent model of similar spirit, an energy functional has been postulated to characterize the behavior of crystalline structures through a density field description. We refer to the phase field crystal model suggested by Elder and Grant \cite{elder2002modeling,elder2004modeling}. Periodicity of the density field is put into the model by energetically stabilizing at least one nonzero wave vector in the ground state. The magnitude of the subsequent periodic density field is interpreted as the averaged probability to find a constituent of the crystal. Short timescales of motion are regarded as averaged over. Because of this, the approach is very effective in modeling processes in crystals on timescales that are long from a molecular dynamics point of view. We refer to such a model as a density field description because the order parameter is related to a mass density rather than a concentration field. 

Kinetic evolution of periodic patterns from a disordered state and corresponding phase diagrams have been analyzed in the framework of the phase field crystal model \cite{elder2004modeling}. Furthermore, the approach has been used to describe the motion of defects in crystalline structures \cite{berry2006diffusive,chan2009molecular}, including avalanche events \cite{chan2010plasticity}. The phase field crystal model has been shown to be a starting point for multiscale simulations of multidomain structures \cite{athreya2007adaptive,provatas2007using}. In addition, the suitability for the characterization of epitaxial growth and crack propagation has been pointed out \cite{elder2002modeling,elder2004modeling}. It has been demonstrated that averaging the trajectories obtained from molecular dynamics simulations indeed leads to density fields that can be captured by the phase field crystal model \cite{tupper2008phase}. The connection to classical density functional theory is of current interest \cite{elder2007phase,teeffelen2009derivation}. 

Both of the two approaches outlined above can be interpreted as describing a sort of phase separation on micro- or mesoscopic length scales. This means that macroscopic coarse-graining of the observed structures, which sets in, for example, in the case of the Cahn-Hilliard model \cite{cahn1958free}, is suppressed. The structures that are obtained are periodic in space. Or, at least, if we think of defect-rich and/or multi-domain structures, the energetically favored ground state is periodic in space. It is our goal in this paper to describe a sort of microphase separation that does not feature this tendency and is nonperiodic in space. Whereas the phase field crystal model has been successful to characterize processes in crystalline materials, our approach may be interesting to model amorphous materials such as polymers. 

In the next section we phenomenologically introduce such a nonlocal energy functional and motivate the concept behind it. We explain how we numerically solve the corresponding kinetic equation in the appendix. After that, in section \ref{textures}, we give an overview on the textures that are obtained from minimizing the energy functional. In particular, we focus on the time evolution of the structures as they develop from the disordered state. At this stage, coarse-graining to macrophase separation still sets in. We therefore present a modification to our model in section \ref{stripes}, which leads to stabilized stripe phases that do not phase separate macroscopically. Some possible applications, predominantly in the field of soft matter systems, are pointed out in section \ref{applications}, before we conclude.

\section{\label{potential}Simple nonlocal energy functional}

\begin{figure}
\includegraphics[width=7.5cm]{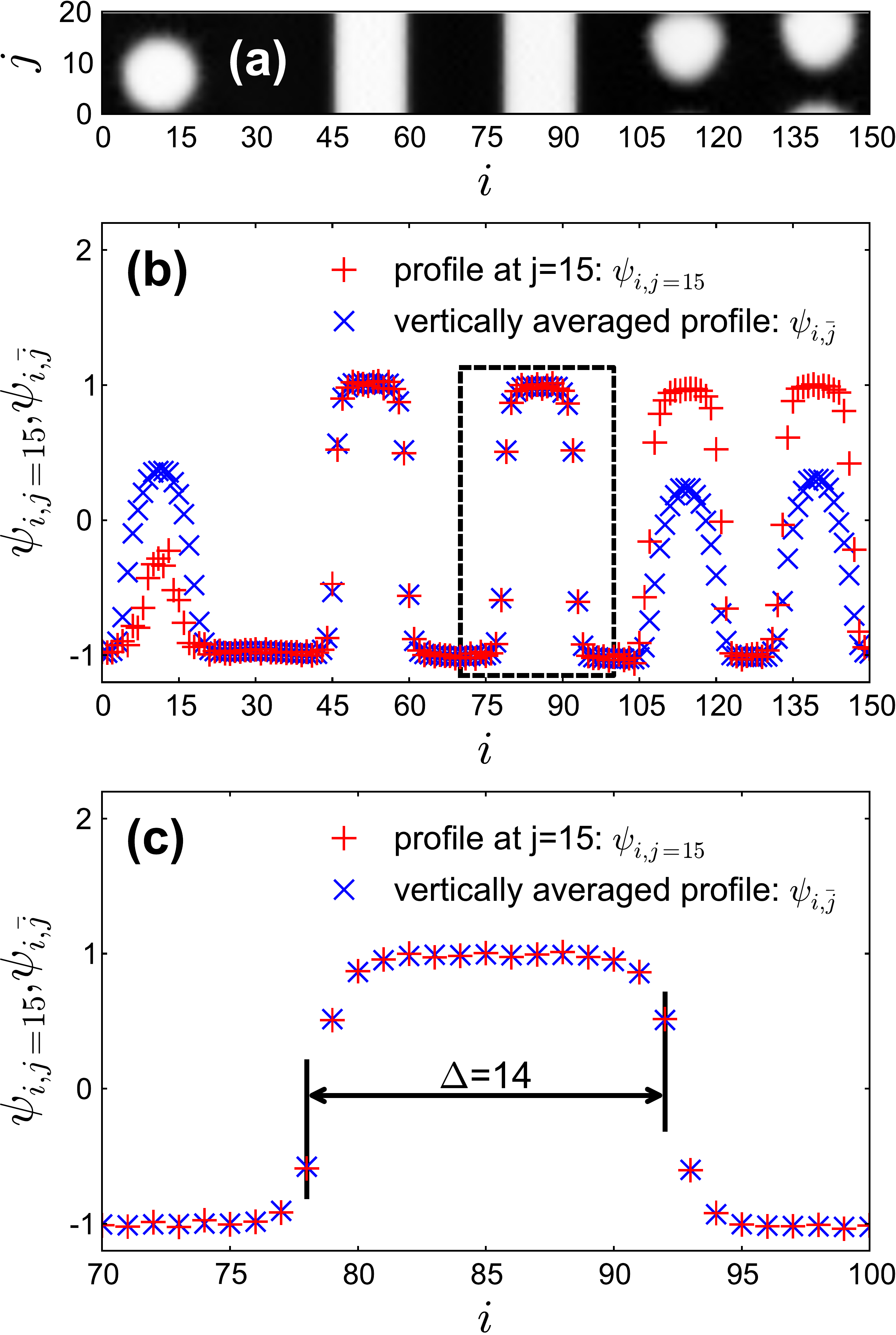}
\caption{(Color online). (a) Textures observed from a numerical iteration of Eqs.~(\ref{F}) and (\ref{kinetic-eq}) forward in time, starting from a homogeneous disordered state. Brightness indicates the value of $\psi$, where black and white correspond to values around $-1$ and $+1$, respectively. The actual calculation domain was of size $1000\times20$. Only a domain of size $150\times20$ is shown. The boundary conditions are periodic. (b) Density profiles $\psi_{i,j=15}$ along a horizontal line at height $j=15$ in Fig.~(a) ($+$ symbols) and $\psi_{i,\bar{j}}$ averaged over the $j$-direction of Fig.~(a) ($\times$ symbols). (c) Zoom into the area in (b) surrounded by the dashed rectangle. Gradients $\nabla\psi$ at the lattice sites pairwisely sum up to zero, as indicated for one pair. The distance between the points of each such pair is $\Delta=14$ lattice distances. (Technical details: $\vartheta=5$, $\alpha=3$, $d=15$, $L=1$, $\bar{\psi}=-0.4$, lattice distance $dx=1$, $800000$ time steps of step size $dt=0.0008$ yielding a time $t=640$).}
\label{streifen}
\end{figure}

As pointed out above, we now introduce a simple energy functional that produces localized nonperiodic structures. The functional is introduced in a phenomenological way. However, we will shortly explain the concept behind the chosen functional form using the illustration in Fig.~\ref{streifen}. After that, we will point out the sort of textures obtained. We will restrict ourselves to two spatial dimensions in this paper. 

Our goal is to use a single scalar order parameter field $\psi(\mathbf{r})$ to describe localized nonperiodic objects, as for example the ones shown in Fig.~\ref{streifen}(a). 
$\psi(\mathbf{r})$ denotes a concentration or density field, similar to the continuum field description on block copolymers \cite{ohta1986equilibrium} and the phase field crystal model \cite{elder2004modeling}. The spheres and stripes shown in Fig.~\ref{streifen}(a) feature a characteristic size $d$ that we externally impose. It corresponds to the diameter and thickness of the spheres and stripes, respectively. In this example, the distances between the objects do not follow a periodic rule. More details are given below. 

Without loss of generality, we identify the black and white regions in Fig.~\ref{streifen}(a) with values $\psi=-1$ and $\psi=+1$, respectively. Fig.~\ref{streifen}(c) shows the density or concentration profile along a horizontal cut through one of the stripes. 

We now consider a single isolated up-gradient, as shown, for example, in the left half of Fig.~\ref{streifen}(c). It would be energetically unstable in periodic approaches like the ones introduced above \cite{ohta1986equilibrium,elder2004modeling} because it is not embedded in an overall periodic structure. Our idea is to balance each such local up-gradient by a down-gradient that is located a distance $d$ away in the direction the gradient is pointing to. This direction is given by the unit vector $\nabla\psi/\|\nabla\psi\|$. In this way, we form localized pairs of gradients that energetically balance each other, having the same magnitude but opposite sign and being separated by an imposed distance $d$ ($\approx\Delta$ in Fig.~\ref{streifen}(c), see below). 

Translating this into a formula, the energy functional $\mathcal{F}$ that we will investigate is based on the energy density 
\begin{equation}\label{F}
F(\mathbf{r}) = \frac{\vartheta}{4}\left(\psi^2|_{\mathbf{r}}-1\right)^2 
    + \alpha\left[ \nabla\psi|_{\mathbf{r}} 
    + \nabla\psi|_{\mathbf{r}+d\frac{\nabla\psi}{\|\nabla\psi\|}}\right]^2.
\end{equation}
Here, $\vartheta$ is a measure for the inverse temperature. The part in square brackets is weighted by a second constant parameter $\alpha$ and contains the impact of the gradient field $\nabla\psi$. Here, the first term refers to $\nabla\psi$ at position $\mathbf{r}$ and is local. The second term, however, refers to the value of $\nabla\psi$ at a position of distance $d$ away from $\mathbf{r}$ in the direction that $\nabla\psi$ points to. This latter term is therefore a nonlocal contribution. Expression (\ref{F}) satisfies the necessary symmetry requirements. To guarantee thermodynamic stability, we must set $\vartheta>0$ and $\alpha>0$. We obtain the energy functional $\mathcal{F}\{\psi(\mathbf{r})\}$ via $\mathcal{F}=\int F(\mathbf{r})\,d^2\hspace{-.7pt}r$. 

The kinetic equation for the time evolution of the conserved field $\psi(\mathbf{r})$ follows from minimizing the energy $\mathcal{F}$, 
\begin{equation}\label{kinetic-eq}
\frac{\partial \psi}{\partial t} = L\, \Delta \frac{\delta \mathcal{F}}{\delta \psi}.
\end{equation}
Mostly in this paper, the parameter $L$ will be set equal to one, $L\equiv 1$, which can always be achieved by rescaling the time. 
On the right-hand side of this equation, the Laplacian reflects the fact that $\psi$ is a conserved quantity. The nonlocal contribution of Eq.~(\ref{F}) is implicitly contained in the functional derivative $\delta\mathcal{F}/\delta\psi$. 

We must remark at this point that, through the form of Eq.~(\ref{kinetic-eq}), we limit our investigations to the regime of relaxation dynamics. Processes on time scales that are shorter than the diffusive one have to be considered as being averaged over. Such processes cannot be resolved explicitly by this approach. As a benefit from this coarse-grained procedure, however, longer time scales are accessible by numerical calculations. The same approach was used in the introduced studies on block copolymer systems \cite{bahiana1990cell,nonomura2001growth,yamada2007interface} and the phase field crystal model \cite{elder2002modeling,elder2004modeling,berry2006diffusive, athreya2007adaptive,tupper2008phase,elder2007phase,teeffelen2009derivation}, or, in a similar spirit, also in classical dynamical density functional theory \cite{marconi1999dynamic,archer2004dynamical,espanol2009derivation,teeffelen2009derivation}. 

Due to the nonlocal contribution, we could not perform the functional derivative on the right-hand side of Eq.~(\ref{kinetic-eq}) analytically. We present our way of numerical evaluation of the functional derivative in the appendix and the supplemental material \cite{supplemental}. The central point in our discretized numerical evaluation is to track the connections between the points $\mathbf{r}$ and $\mathbf{r}+d\frac{\nabla\psi}{\|\nabla\psi\|}$. Due to the discrete nature of the numerical results, we will replace the continuous position vector $\mathbf{r}$ by indices $i$ and $j$. 

In the current form, Eq.~(\ref{F}) contains three parameters $\vartheta$, $\alpha$, and the distance $d$. It follows directly that $\alpha$ can be scaled out. For practical reasons, we set $\alpha\equiv 3$ in all our numerical considerations. We further note that in the limit $d\rightarrow 0$, our expression for $F$ leads to the famous Cahn-Hilliard equation \cite{cahn1958free}. 

A first illustrative example created from our nonlocal energy functional is actually given by Fig.~\ref{streifen}. It shows the result of an iteration of Eqs.~(\ref{F}) and (\ref{kinetic-eq}) forward in time on a rectangular domain of large aspect ratio and periodic boundary conditions. The axis labels $i$ and $j$ index the grid sites of the square calculation grid. We initialized the field $\psi$ as $\psi_{ij}=\bar{\psi}+\eta_{ij}$, with $\bar{\psi}=-0.4$ and $\eta_{ij}$ a random field of zero mean, Gaussian distribution, and small amplitude. Sphere and stripe domains are obtained. Their diameter follows from the value of the parameter $d$, which we set to $d=15$. 

We display in Fig.~\ref{streifen}(b) the density profile $\psi_{i,j=15}$ on a horizontal line at $j=15$, and a vertically averaged profile $\psi_{i,\bar{j}}=\sum_{j=0}^{20}\psi_{ij}/21$. The values of the two profile functions are equal for the stripes and differ for the spheres. 

Fig.~\ref{streifen}(c) is a zoom into the area of Fig.~\ref{streifen}(b) that is marked by the dashed rectangle and depicts the concept behind Eq.~(\ref{F}). The first part with the coefficient $\vartheta$ enforces concentrations of $\psi=\pm 1$ for $\vartheta>0$. All other values increase this contribution to the energy of the system. The second part with the coefficient $\alpha$ leads to an energetic penalty for gradients $\nabla\psi$, {\em unless} there is a gradient of opposite sign, but same magnitude, located at a distance $d$ away in the direction that $\nabla\psi$ points to. Indeed, this pairwise annihilation of gradients occurs as indicated for one pair in Fig.~\ref{streifen}(c). (The distance is $\Delta=14$ and not $d=15$ because of an integer cast of the term $d\frac{\nabla\psi}{\|\nabla\psi\|}$ in our numerical calculation to index the discrete lattice sites). In this way, no energetic penalty arises from the gradient terms in Eq.~(\ref{F}). 
 
In summary, we do not confine ourselves to local expressions in the gradient fields. Such local expressions have the tendency to enforce periodicity throughout the whole space. This becomes clear from the Fourier transforms of the corresponding energy densities. Usually, a certain band of wave numbers is preferred homogeneously over the whole space. Our goal is to generate objects that are localized in space and not embedded in an overall tendency of periodicity. This is different from multigrain structures or defect-rich textures since in the latter cases a periodic order is still energetically favored. In other words, we wish to enforce phase separation on a mesoscopic length scale that leads to amorphous textures as observed in non-crystallized polymeric systems, for example, or localized objects such as vesicles floating in a fluid environment, using one scalar order parameter field.

\section{\label{textures}Time evolution of the textures}

In the following, we describe the kinetic evolution of the patterns that are generated by Eqs.~(\ref{F}) and (\ref{kinetic-eq}). Since we compare to results obtained for the Cahn-Hilliard equation, we refer to the order parameter $\psi$ as a concentration field. We will start our numerical considerations in all cases from a spatially homogeneous concentration that is modulated only by a random field of zero mean, Gaussian distribution, and small amplitude. The numerical calculations were performed in the way pointed out in the appendix and the supplemental material \cite{supplemental}. 

\begin{figure*}
\includegraphics[width=17.cm]{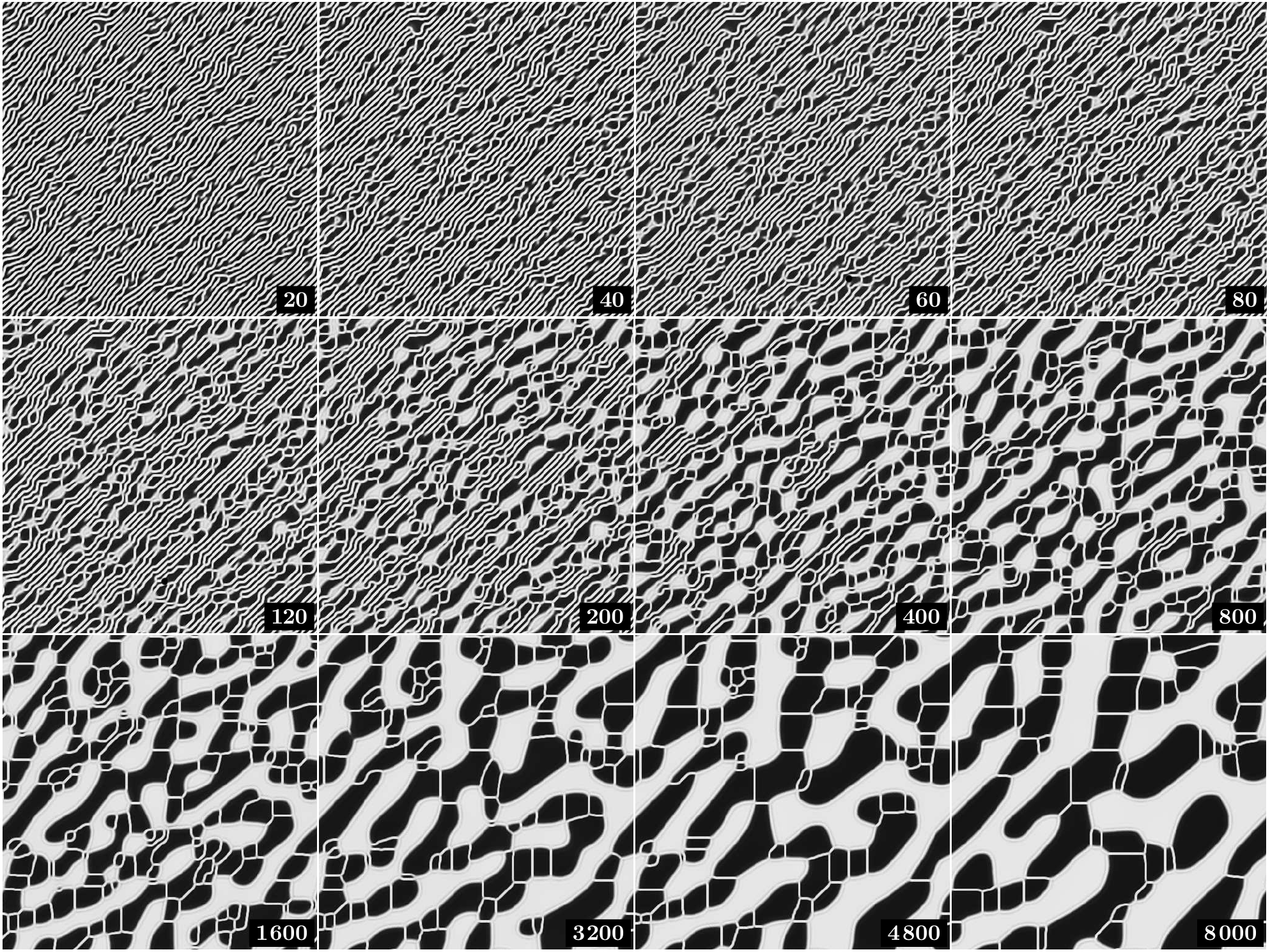}
\caption{Time series of the kinetic evolution described by Eqs.~(\ref{F}) and (\ref{kinetic-eq}), with a mean concentration $\bar{\psi}=0$. The gray scale represents the concentration values, where white and black correspond to the maximum and minimum concentration at each time, respectively. Parameter values were set to $\vartheta=5$, $\alpha=3$, $d=5$, the grid size to $512\times512$ sites of distance $dx=1$. Time steps were of size $dt=0.0008$, the times of taking the pictures are indicated in their bottom right corners.}
\label{timeseries00}
\end{figure*}

\begin{figure*}
\includegraphics[width=17.cm]{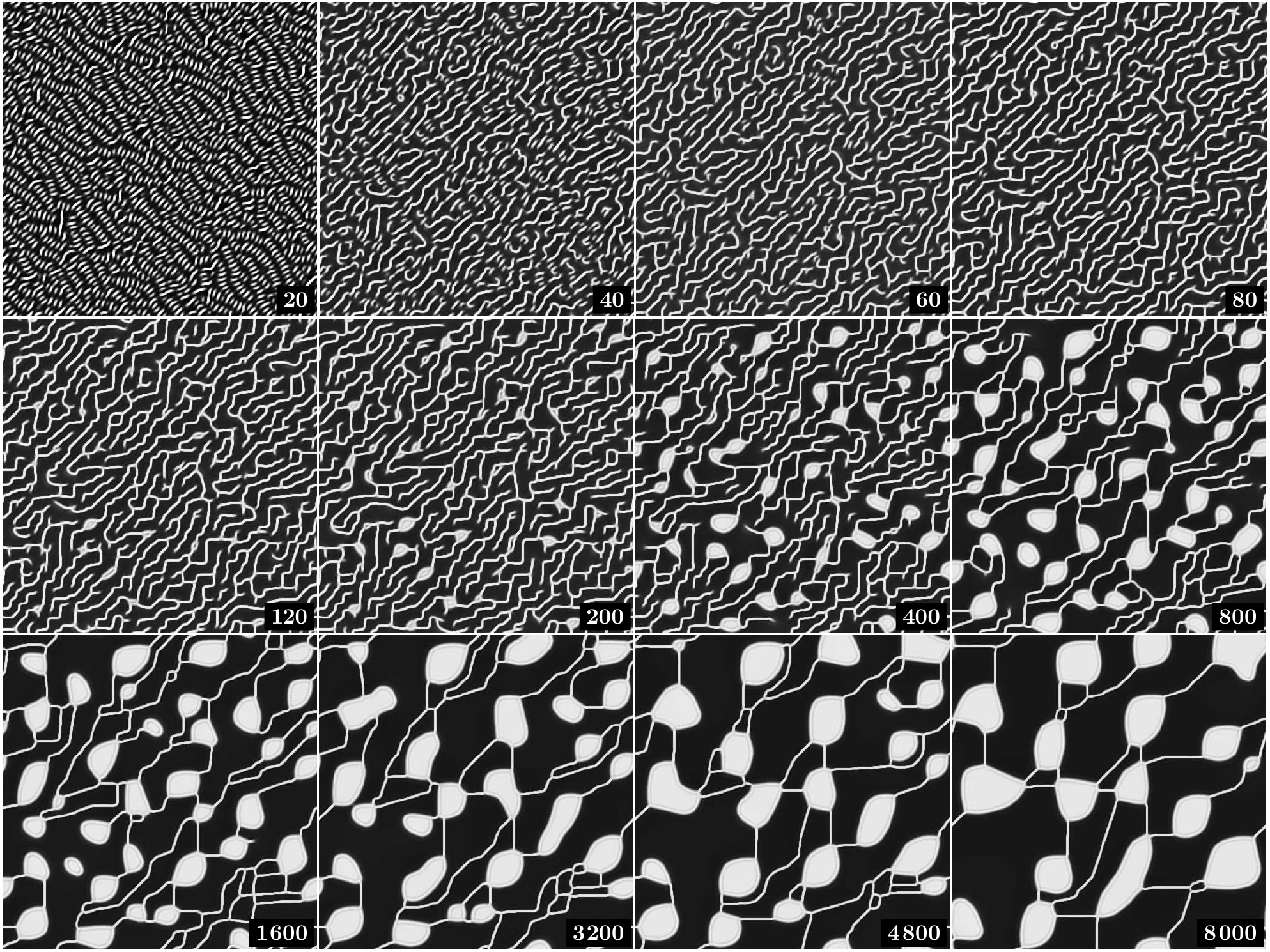}
\caption{Time series of the kinetic evolution described by Eqs.~(\ref{F}) and (\ref{kinetic-eq}), with a mean concentration $\bar{\psi}=-0.4$. The other settings were the same as described in the caption of Fig.~\ref{timeseries00}.}
\label{timeseries04}
\end{figure*}

We show in Fig.~\ref{timeseries00} a time series of phase separation for a mean concentration $\bar{\psi}=0$. Furthermore, we chose $\vartheta=5$, $\alpha=3$, $d=5$, and a grid size of $512\times512$ lattice sites. The brighter the color in the pictures, the higher the concentration (for each picture the concentration values were normalized by their maximum value). Fig.~\ref{timeseries04} represents the analogous time series for a mean concentration $\bar{\psi}=-0.4$. 

\begin{figure}
\includegraphics[width=8.5cm]{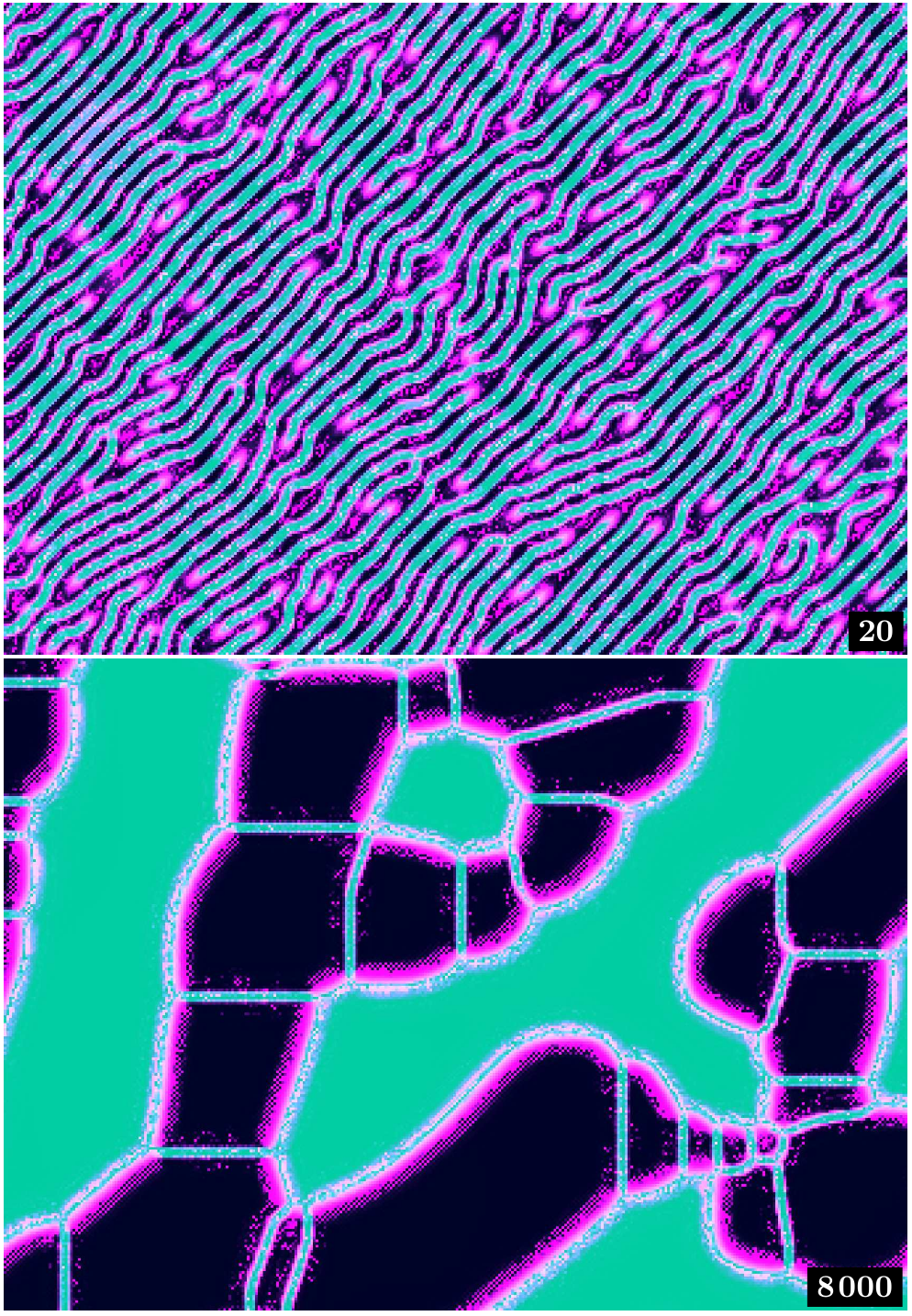}
\caption{(Color online). Energy density $F$ calculated according to Eq.~(\ref{F}) for two of the snapshots presented in Fig.~\ref{timeseries00}: the ones taken at time $t=20$ and at time $t=8000$. The reddish color marks regions of high energy density. Clearly, the areas around defects and at the surface of the blobs stand out. The pictures show averages over $100$ iterations after the given time to smooth the fluctuations that arise from the discrete nature of the numerical implementation. Only a domain of $170\times128$ lattice sites is shown.}
\label{energdens}
\end{figure}

First, thread-like stripe structures form. Shorter threads connect to longer threads. The reason for these steps of forming longer objects is the higher energy density at the ends of the threads: there is no concentration gradient within the threads that could balance the longitudinal concentration gradient at the end points. This increases the local energy density (compare Eq.~(\ref{F}) and Fig.~\ref{streifen}). The upper part of Fig.~\ref{energdens} highlights the spatial distribution of the energy density at an early stage of pattern evolution for the case $\bar{\psi}=0$. 
Ends of the threads clearly stand out. We further investigate this initial process of forming and connecting to longer threads in the next section. 

After that, another process of coarse-graining sets in, which leads to macroscopic phase separation. Blobs start to form at defects in the thread structure and then preferentially grow and coarse-grain along the thread connections. The smaller the mean concentration $\bar{\psi}$, the less connected are the blobs by threads during the process of coarse-graining. As we will show below, this coarse-graining process is similar to the one observed for the Cahn-Hilliard model. 

Energetically, we can explain this coarse-graining to a macroscopic phase separation in the following way. The stripe patterns as they evolve from Eqs.~(\ref{F}) and (\ref{kinetic-eq}) correspond to a metastable state. This is true even if the high-energetic end points of the threads are excluded from our considerations. Due to pairwise annihilation of the gradients at the boundaries of the stripes, the term with coefficient $\alpha$ can be minimized in Eq.~(\ref{F}). However, the values at the boundaries of the stripes deviate from $\psi=\pm1$ (compare Fig.~\ref{streifen}), which increases the $\vartheta$-contribution. 

For the blobs, the situation is just the other way around. The gradients at the surfaces cannot be annihilated, so the $\alpha$-term in Eq.~(\ref{F}) increases in magnitude. This leads to the high energy density at the surfaces depicted in the lower part of Fig.~\ref{energdens}. Yet, this scenario is energetically beneficial. In most of the interior of the blobs we find $\psi\approx1$, which implies $F\approx0$. Altogether, this decreases the total energy of the system due to the high number ratio of interior to surface points. 

To characterize the coarse-graining process more quantitatively, we determined the structure function
\begin{equation}\label{Sveck}
S(\mathbf{k},t) = \frac{1}{N}\left\langle \sum_{\mathbf{r}}\sum_{\mathbf{r'}}
  e^{-i\mathbf{k}\cdot\mathbf{r}}
  \left[ \psi(\mathbf{r}+\mathbf{r'},t)\psi(\mathbf{r'},t)-\bar{\psi}^2 \right]
  \right\rangle.
\end{equation}
Here, $N$ is the total number of lattice sites. The sums run over all lattice sites ($\mathbf{r}$ and $\mathbf{r'}$ correspond to index pairs $(i,j)$ and $(l,m)$, respectively, that index the lattice sites). Likewise, the vector $\mathbf{k}$ refers to discrete lattice sites of the corresponding discrete reciprocal lattice. $\langle\dots\rangle$ denotes an ensemble average. In our case, we average over three independent numerical runs that only differ in the initial conditions. More precisely, the three numerical calculations start from different spatial realizations of the narrow Gaussian distribution of concentrations around the spatially homogeneous state. 

From $S(\mathbf{k},t)$, we calculate the circularly averaged structure function 
\begin{equation}\label{Sscalk}
S(k,t)=\frac{1}{\#}\sum_{k'\geq k}^{k'< \,k+dk}S(\mathbf{k'},t). 
\end{equation}
$\#$ stands for the number of lattice sites $\mathbf{k'}$ that satisfy the requirement $k\leq k'<k+dk$, where we set $dk=2$. Despite the anisotropy in our numerical results, the circularly averaged $S(k,t)$ is suitable for what we demonstrate below. 

\begin{figure}
\includegraphics[width=8.5cm]{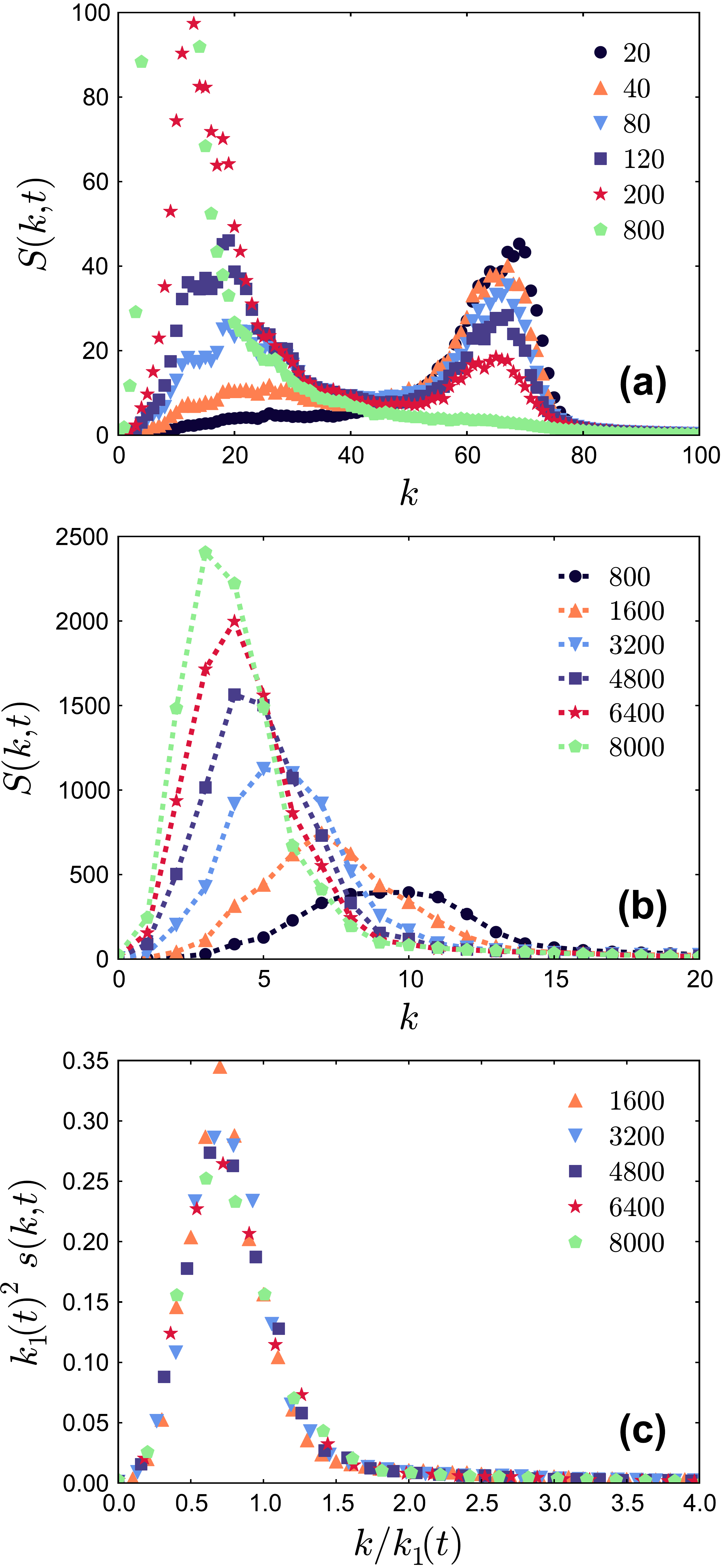}
\caption{(Color online). Circularly averaged structure function $S(k,t)$ for a texture evolved from Eqs.~(\ref{F}) and (\ref{kinetic-eq}) for $\bar{\psi}=0$. The curves correspond to times $t$ as indicated on the upper right of each plot. See Fig.~\ref{timeseries00} for a real space representation and the parameter values. (a) Temporal evolution of $S(k,t)$ during the early stage. The decreasing first order peak at $k\approx 65$ represents the thread-like stripe pattern, the increasing peak at lower wave numbers develops due to the macroscopic coarse-graining. Only data $S(k,t)<100$ are shown. (b) Evolution of $S(k,t)$ at the later stage of coarse-graining. As the blobs grow in size, the peak shifts to lower $k$-values. The dashed lines are guides to the eye only. (c) Data collapse for the later stage of coarse-graining (see text and Ref.~\cite{zhu1999coarsening}).}
\label{strucfunc}
\end{figure}

In Fig.~\ref{strucfunc}, we plot the structure function $S(k,t)$ that corresponds to the case of $\bar{\psi}=0$ and $d=5$ (compare the time series in Fig.~\ref{timeseries00}). 
Fig.~\ref{strucfunc}(a) shows $S(k,t)$ at the early times, when the stripe texture leads to a peak around $k\approx 65$ (we do not display the higher order peaks). This peak indicates a periodic length of about $512/65\approx8$ lattice distances $dx$ of the thread-like stripe structures. Comparing with the first row of pictures in Fig.~\ref{timeseries00}, we confirm this length scale: the threads have a thickness of $d=5$, and, where they are compactly packed, are separated by roughly a distance $d/2$, which gives $d+d/2\approx8$. As time proceeds, this peak becomes smaller until it has vanished at $t\approx800$. 

As the peak at $k\approx 65$ decreases, another peak at smaller $k$-values evolves. This peak corresponds to the onset of the second stage of coarse-graining. Fig.~\ref{strucfunc}(b) illustrates the development of $S(k,t)$ at times longer than $t=800$. The peak shifts to smaller $k$-values, indicating a growth of the characteristic size of the observed structures. In particular, this is reflected by the last row of pictures in Fig.~\ref{timeseries00} and indicates a macroscopic phase separation. 

To discuss the latter statement in more detail, we demonstrate that the same scaling behavior is found as for the later stages of coarse-graining in the Cahn-Hilliard model \cite{cahn1958free,zhu1999coarsening}. We follow Ref.~\cite{zhu1999coarsening} and define a normalized structure function
\begin{equation}
s(k,t)=\frac{S(k,t)}{N\left[ \langle\psi^2\rangle - \langle\psi\rangle^2 \right]}, 
\end{equation}
as well as a characteristic wave number
\begin{equation}
k_1(t) = \frac{\sum_{k}k\, s(k,t)}{\sum_{k}s(k,t)}.
\end{equation}
When we plot $k_1(t)^2s(k,t)$ as a function of $k/k_1(t)$, the data for the later times collapse as displayed by Fig.~\ref{strucfunc}(c). 

\begin{figure}
\includegraphics[width=8.5cm]{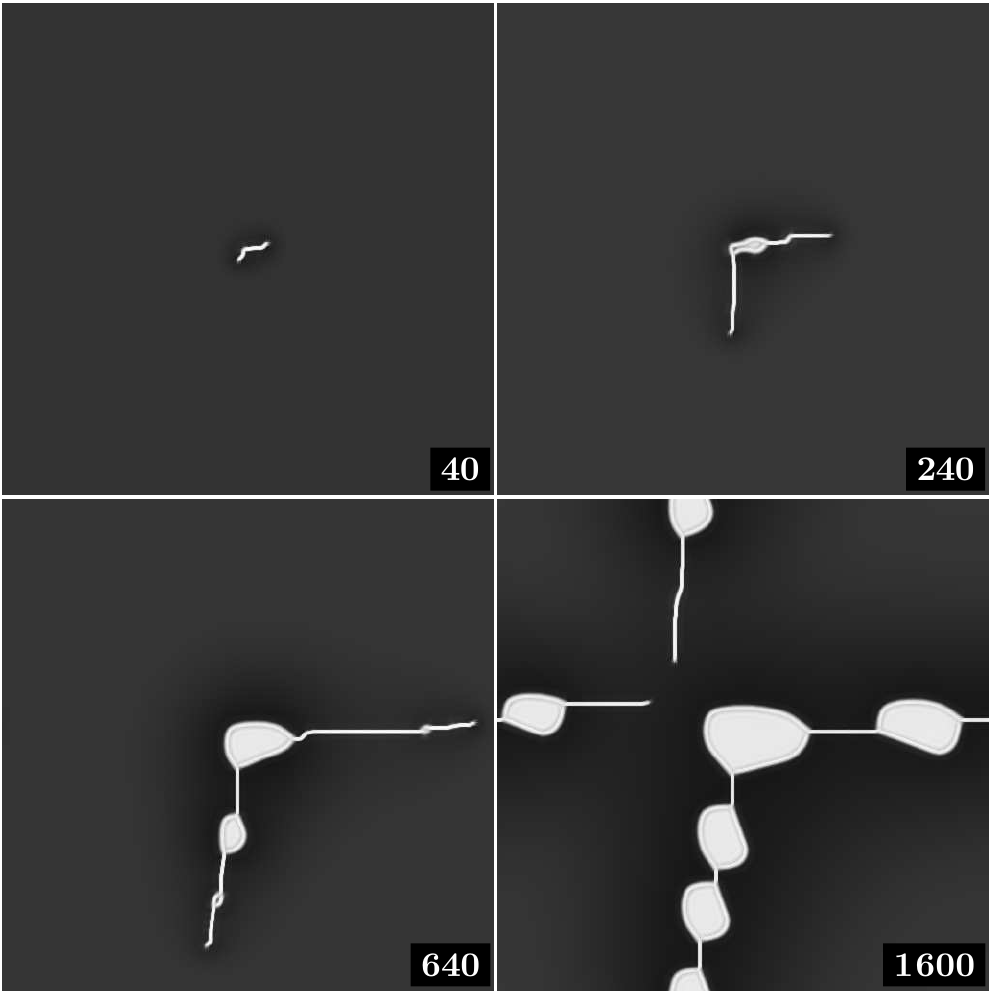}
\caption{Time series of the kinetic evolution described by Eqs.~(\ref{F}) and (\ref{kinetic-eq}), with a mean concentration $\bar{\psi}=-0.7$. The other settings were again the same as described in the caption of Fig.~\ref{timeseries00}. However, a nucleation core of $5\times5$ lattice points and $\psi=0.5$ was initially provided in the center of the calculation grid.}
\label{timeseries07}
\end{figure}

After the analysis of this scaling behavior, we further increased the magnitude of the mean concentration $|\bar{\psi}|$. For values $|\bar{\psi}|\geq0.6$, we found that the patterns did not form spontaneously. This can also be understood from a comparison with the Cahn-Hilliard equation. In the Cahn-Hilliard model, an initially homogeneous state only becomes {\em linearly} unstable for mean concentrations $|\bar{\psi}|\leq1/\sqrt{3}\approx0.577$. We therefore provide a ``condensation nucleus'' at these concentrations $|\bar{\psi}|\geq0.6$. Fig.~\ref{timeseries07} shows an example, where we set $\bar{\psi}=-0.7$. 
The other parameters were chosen the same as in Figs.~\ref{timeseries00} and \ref{timeseries04}. However, at $t=0$, a spot of $5\times5$ lattice points in the center of the figure was set to $\psi=0.5$. Indeed, we then observe a process of growth after providing the initiating nucleation core. Coarse-graining starts at the most unstable points in the threads of Fig.~\ref{timeseries07}. 

\begin{figure}
\includegraphics[width=8.5cm]{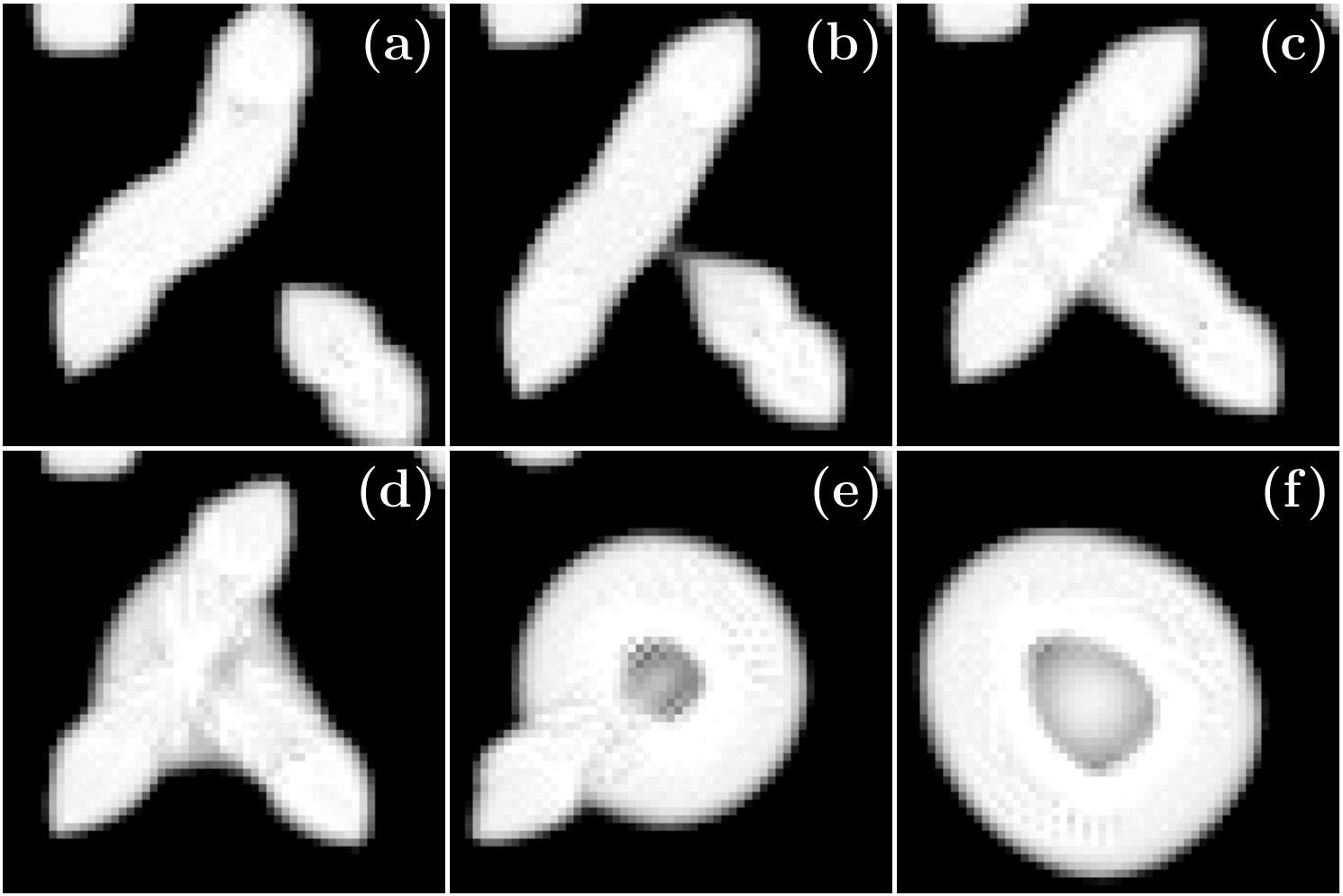}
\caption{Snapshots of $57\times57$ lattice sites from a numerical integration of Eqs.~(\ref{F}) and (\ref{kinetic-eq}), with a mean concentration $\bar{\psi}=-0.4$. We set the parameter values to $\vartheta=5$, $\alpha=3$, $d=15$, the total grid size was $512\times512$ sites of distance $dx=1$, and time steps were of size $dt=0.0008$. The times $t$ that the different pictures correspond to are (a) $5760$, (b) $6400$, (c) $6480$, (d) $6760$, (e) $7200$, and (f) $8000$.}
\label{bildausschnitte04d15}
\end{figure}

Finally, we display in more detail an example of how defects can form, which then initiates the coarse-graining. The pictures in Fig.~\ref{bildausschnitte04d15} show snapshots from a numerical calculation of a mean concentration $\bar{\psi}=-0.4$ and a stripe thickness $d=15$. The series starts from two threads (a), the energy density at the ends of which is increased as noted above. First, the ``reactive'' end of the shorter thread couples into the side of the other thread (b). The resulting structure (c) is frustrated since the requirement of pairwise annihilation of the gradients cannot be satisfied around the connection point. This higher energy density around the crosslinking point initiates the coarse-graining (d). A growing blob incorporates the remaining thread texture (e). Finally, we end up with a spherical object (f) that looks reminiscent of a small vesicle. 

We shortly remark at the end of this section, that the preferentially diagonal, horizontal, and vertical orientation of the threads is due to our discretization scheme of the gradient field as described in the appendix. It reflects the underlying square lattice that acts similar to an external orienting field. Varying the value of the parameter $d$, we checked that the coarse-graining is not due to the interplay with the discrete numerical grid. Still, the amorphous and localized character of the structures is evident, in particular for lower mean concentrations $\bar{\psi}$ (Figs.~\ref{timeseries04} and \ref{timeseries07}). We checked the numerical scheme by varying the size of the time steps $dt$, the distance of the lattice sites $dx$, the characteristic length $d$ (also to non-integer values), as well as the discretization scheme of the Laplacian in Eq.~(\ref{kinetic-eq}).

\section{\label{stripes}Stabilized stripe phases}

To make our model more attractive for the description of amorphous and localized structures, we need to stabilize the stripe state against macroscopic coarse-graining. 
This becomes clear when we consider surfactant molecules as an example. The white regions in Figs.~\ref{timeseries00} and \ref{timeseries04} are identified with the hydrophobic components, the black regions with the hydrophilic components. Then macroscopic phase separation is not allowed. The surfactant molecules cannot simply chemically split their hydrophilic from their hydrophobic parts. At the same time, each gradient in $\psi$ is associated with an accumulation of the hydrophilic component on the up-side and the hydrophobic component on its down-side. 

In the following minimal approach, we only consider the accumulation on the down-side of each gradient. We do this by adding an additional contribution $F_{\beta}$ to the energy density in Eq.~(\ref{F}),
\begin{equation}\label{Fbeta}
F_{\beta}(\mathbf{r})=\beta\left[ \psi|_{\mathbf{r}-b\frac{\nabla\psi}{\|\nabla\psi\|}} +1 \right]^2.
\end{equation}
Due to this contribution, each gradient $\nabla\psi$ favors a value of $\psi\approx-1$ in its downhill direction. More precisely, this term leads to an energetic penalty at the position $\mathbf{r}$, if the value of $\psi$ deviates from $\psi=-1$ at a distance $b$ away in the direction $-\frac{\nabla\psi}{\|\nabla\psi\|}|_{\mathbf{r}}$. The interplay with the $\alpha$-contribution in Eq.~(\ref{F}) energetically disfavors macroscopic coarse-graining. 

We can understand the latter statement by looking at the fine structure of the blob boundaries in Figs.~\ref{timeseries00}, \ref{timeseries04}, \ref{timeseries07}, and \ref{bildausschnitte04d15}. The $\alpha$-term in Eq.~(\ref{F}) enforces the formation of concentration shells around the blob, which corresponds to the scheme depicted by Fig.~\ref{streifen}. It leads to concentration gradients at the inner shell boundary pointing into the inward direction. However, the inside concentration of the blob is $\psi\approx+1$, which leads to a high energetic penalty due to the $\beta$-contribution in Eq.~(\ref{Fbeta}). Therefore, macroscopic coarse-graining is energetically inhibited. Details of the numerical implementation of the $\beta$-term are given at the end of the appendix and in the supplemental material \cite{supplemental}. 

Figs.~\ref{timeseries00beta10} and \ref{timeseries04beta10} illustrate the stability of the stripe patterns when Eq.~(\ref{Fbeta}) is included. The mean concentrations were given by $\bar{\psi}=0$ and $\bar{\psi}=-0.4$, respectively. We set the other parameter values to $\vartheta=5$, $\alpha=3$, $d=5$, $\beta=1$, and $b=3$. Again, the grid size was $512\times512$ lattice points, and the length of the numerical iteration was the same as the one leading to Figs.~\ref{timeseries00} and \ref{timeseries04}. 
\begin{figure}
\includegraphics[width=8.5cm]{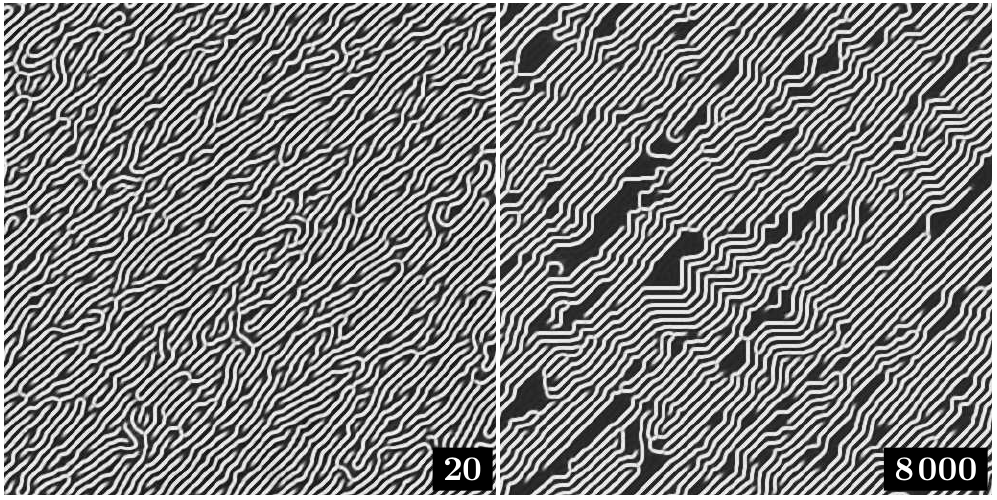}
\caption{Stability of the stripe state during a kinetic evolution described by Eqs.~(\ref{F}) and (\ref{kinetic-eq}), supplemented by Eq.~(\ref{Fbeta}), for a mean concentration $\bar{\psi}=0$. In contrast to Fig.~\ref{timeseries00}, no macroscopic coarse-graining occurs. Parameter values were set to $\vartheta=5$, $\alpha=3$, $d=5$, $\beta=1$, and $b=3$. The grid size was $512\times512$ lattice points of distance $dx=1$, and time steps were of length $dt=0.0008$.}
\label{timeseries00beta10}
\end{figure}
\begin{figure}
\includegraphics[width=8.5cm]{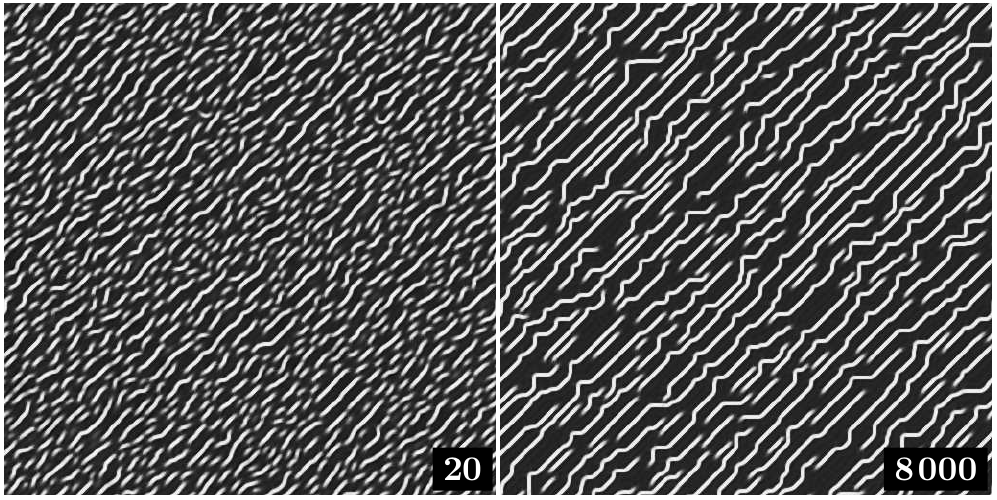}
\caption{Stability of the stripe state during a kinetic evolution described by Eqs.~(\ref{F}) and (\ref{kinetic-eq}), supplemented by Eq.~(\ref{Fbeta}), for a mean concentration $\bar{\psi}=-0.4$. No macroscopic coarse-graining occurs, in contrast to Fig.~\ref{timeseries04}. Parameter values as given in the caption of Fig.~\ref{timeseries00beta10}.}
\label{timeseries04beta10}
\end{figure}

A macroscopic phase separation via the blob formation does not occur, despite the presence of defects and crosslinking points. The only sort of coarse-graining is the formation of longer threads from shorter ones. Two different realizations of this process are observed. On the one hand, mainly short threads can diffuse into other threads. This happens predominantly at the early stage of coarse-graining. It requires a nearby thread that can grow on the cost of the dissolving thread. On the other hand, the ``reactive'' ends of two different threads can combine to form one resulting long thread. If such a ``reactive end'' attacks the inner part of another thread similarly to the process depicted in Fig.~\ref{bildausschnitte04d15}(b), crosslinking points are generated. The latter hardly occurs at lower concentrations (compare Figs.~\ref{timeseries00beta10} and \ref{timeseries04beta10} at time $t=8000$). We have to note that the influence of the underlying square grid becomes rather obvious at the later numerical times. 

In order to describe the process of coarse-graining to the longer threads more quantitatively, we analyzed the kinetics of the end points. We refer to these points as defects in the following. There are several qualitatively different ways defects can evolve. Defects can appear together with the threads when the latter form from the homogeneous disordered state. This process mainly takes place at the very early stage of our numerical calculations and will not be considered. Apart from that, defects can appear when crosslinking points disconnect (inversion of the process shown in Fig.~\ref{bildausschnitte04d15}(a)-(c)). Two defects can unite to one defect when threads become very short and the end points of the same thread meet. We have checked that this process does not play a major role in what follows. Defects further vanish by forming crosslinking points (in analogy to Fig.~\ref{bildausschnitte04d15}(a)-(c)), or by connection of two threads to one at the end points. The latter process dominates particularly at lower mean concentrations $\bar{\psi}$. 

\begin{figure}
\includegraphics[width=8.5cm]{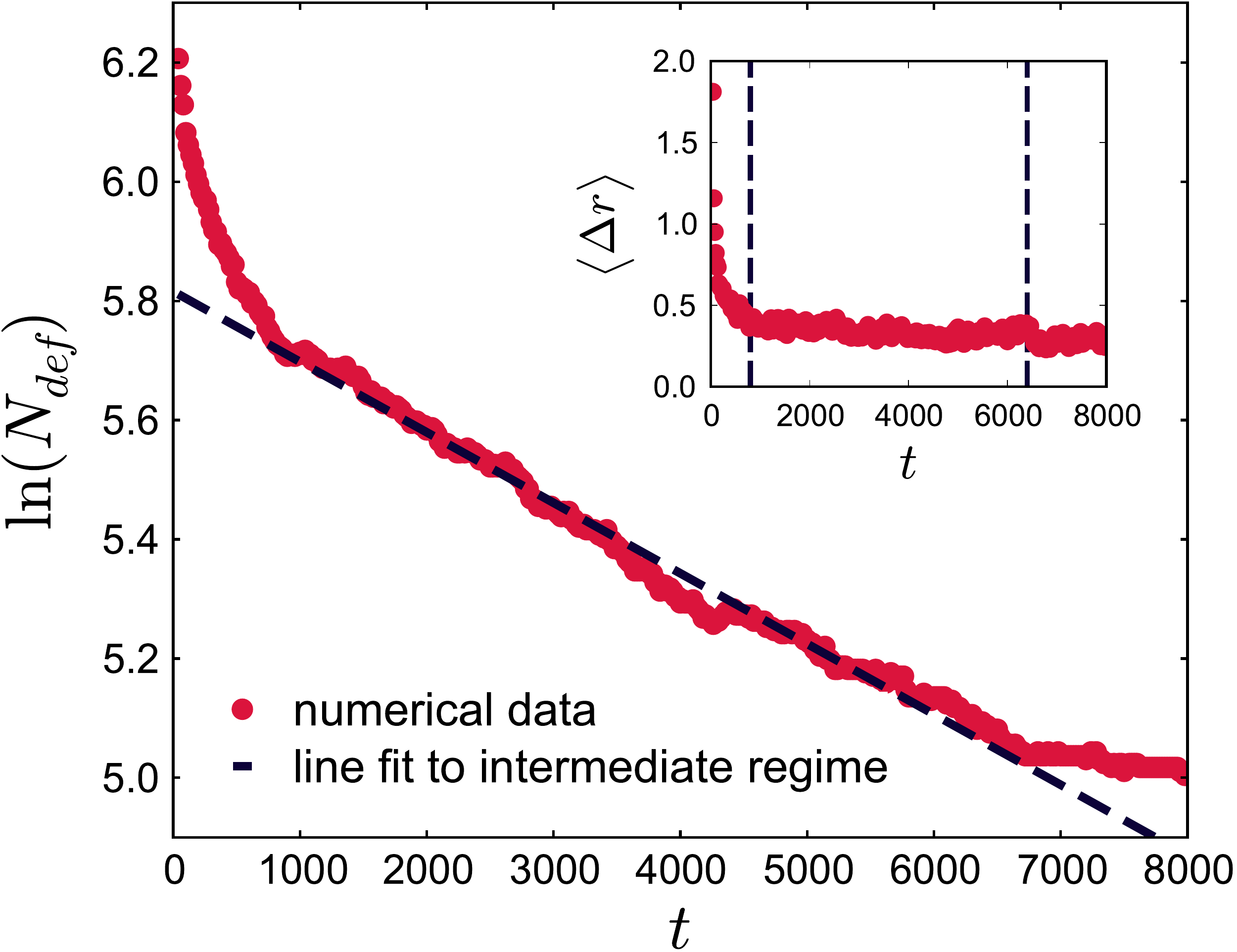}
\caption{(Color online). Number of defects (end points of the threads) $N_{de\!f}$ in the stripe patterns as a function of numerical calculation time $t$ in a semilogarithmic plot. As marked by the straight line, the defect number decays exponentially in a broad regime from $t=800$ to $t=6400$. The inset shows the absolute displacement distance of the defects $\Delta r$ per time interval $\Delta t=20$ and averaged over the whole numerical sample. Dashed vertical lines mark the interval used for the line fit. Parameter values as given in the caption of Fig.~\ref{timeseries00beta10}}
\label{defektzerfall}
\end{figure}

Fig.~\ref{defektzerfall} shows the number of the defects $N_{de\!f}$ on a calculation grid of $512\times512$ lattice sites as a function of time. 
The mean concentration was $\bar{\psi}=0$, and the other parameter values were chosen as mentioned in the caption of Fig.~\ref{timeseries00beta10}. Directly after the initial quench from the homogeneous state to a temperature given by $\vartheta=5$, the number of defects decreases rapidly. As given by the straight line in the semilogarithmic plot, we then found a broad regime from about $t=800$ to $t=6400$, where the defect number decreases exponentially. At later times the annihilation of defects seems to saturate. Apart from that, we tracked the motion of the defects. In the regime of exponential decay, the average distance of motion of the defects $\langle\Delta r\rangle$ was relatively constant with about $0.34$ lattice sites per time interval $\Delta t=20$. The corresponding standard deviation was less than $0.04$ lattice sites. We have plotted the time dependence of $\langle\Delta r\rangle$ in the inset of Fig.~\ref{defektzerfall}. There, the dashed vertical lines mark the interval that we used for the line fit in the main part of the plot.

We performed three independent numerical runs with different initial spatial Gaussian fluctuations around the homogeneous state. Out of these three runs, we chose the one where the defect number saturated the latest. The other two samples with earlier defect number saturation did not reveal a well-defined intermediate exponential regime. 

\begin{figure}
\includegraphics[width=8.5cm]{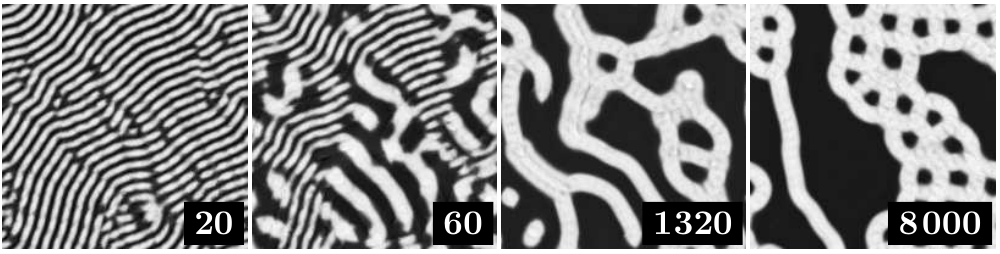}
\caption{Time series of the kinetic evolution of Eqs.~(\ref{F}), (\ref{kinetic-eq}), and (\ref{Fbeta}), with a mean concentration $\bar{\psi}=0$. Parameter values were set to $\vartheta=8$, $d=15$, $\beta=0.8$, and $b=10$, the grid size was $200\times200$ lattice sites of distance $dx=1$, and time steps were of size $dt=0.0008$. First an interlaced structure of thinner threads form. These become unstable and partially coarse-grain to threads of thickness $d$. Then parts of the threads grow thicker than $d$, and finally a bubble-like structure of homogeneous bubble size and wall thickness $d$ emerges.}
\label{timeseries00theta8beta08d15b10}
\end{figure}

At the end of this section we point out that variations of the parameters $\vartheta$, $d$, $\beta$, and $b$ allow for further, qualitatively different structures. We give one example in the timeseries of Fig.~\ref{timeseries00theta8beta08d15b10}. 
There, the mean concentration was set to $\bar{\psi}=0$, and the other parameters were chosen as $\vartheta=8$, $d=15$, $\beta=0.8$, and $b=10$. For higher thicknesses $d$, we usually observe that an interlaced structure of thinner stripes forms from the spatially homogeneous state (first picture of Fig.~\ref{timeseries00theta8beta08d15b10}). These stripes then become unstable with respect to the formation of the threads of thickness $d$. In the case of the current set of parameters, the threads then partially thicken to threads of a cross section higher than $d$ (third picture of Fig.~\ref{timeseries00theta8beta08d15b10}). Finally, the latter transform into a foam-like texture with walls of thickness $d$ and uniform bubble size. The macroscopic phase separation and coarse-graining into blobs as observed in Figs.~\ref{timeseries00}, \ref{timeseries04}, and \ref{timeseries07} does not occur for this set of parameters.

\section{\label{applications}Possible applications}

In this section we discuss two possible applications of our energy density (\ref{F}) and (\ref{Fbeta}) to model systems in the field of soft condensed matter. In the first case, $\psi(\mathbf{r})$ is interpreted as a density field, which is therefore of the same spirit as is the phase field crystal model \cite{elder2004modeling} for periodic systems. In the second part, $\psi(\mathbf{r})$ should rather be interpreted as a concentration field. This can be seen in analogy to the concentration field description of block copolymer systems \cite{ohta1986equilibrium} in the periodic case. 

From the density field point of view, as already mentioned in the introduction, it appears natural to use Eqs.~(\ref{F}) and (\ref{Fbeta}) for a density field characterization of polymeric systems. $\psi(\mathbf{r})$ is then connected to the time averaged probability of finding a segment of a polymer chain at position $\mathbf{r}$, in analogy to the phase field crystal approach \cite{tupper2008phase}. Our model then describes polymeric systems on the length scale of a segment and on diffusive time scales. 

Two steps on this way to a realistic microscopic continuum characterization of polymer systems seem to be important. First, we need to work out a more isotropic discretization scheme than the one presented in the appendix. This is necessary to decouple the orientation of the threads from the orientation of the calculation grid. Second, several topological properties of polymer systems can only arise in three dimensional space. This concerns topological entanglements that are not possible in two spatial dimensions. Special, yet illustrative, examples include mixtures of ring and linear polymers, where the rings can be threaded by the linear chains \cite{klein1986dynamics,mills1987diffusion}, and slide-ring gels \cite{okumura2001polyrotaxane,karino2005sans,deGennes1999sliding}. It will therefore be interesting to generalize our approach to three dimensional space. 

These points are important, e.g., for rheological considerations. Here, we leave them for future work. Instead, we now concentrate on the kinetics of the reaction process illustrated for example in Fig.~\ref{timeseries04beta10}. We note that the process of phase separation and formation of the stabilized stripes reflects several characteristics of step-growth polymerization reactions \cite{munk1989introduction,cowie1991polymers}. In these reactions, all reactants/monomers can participate in the process from the beginning and are quickly incorporated into short chains. The chains can grow at both ends, and short chains can connect to longer ones. This means that the chain length continuously grows with increasing reaction time. Long reaction times are necessary to obtain high degrees of polymerization (long chains). All of these features are reflected by the process referred to in Fig.~\ref{timeseries04beta10}. 

As common in the literature, we now denote by $x$ the number of units/monomers that a chain consists of. $N_0$ gives the total number of units/monomers in the sample, whereas $N$ denotes the total number of molecules of all sizes at a specific reaction time. Then the expression $p=(N_0-N)/N_0$ describes the extent of reaction at that specific reaction time. As shown by Flory, the number $N_x$ of molecules that consist of $x$ units/monomers is given by \cite{flory1936molecular,flory1953principles,cowie1991polymers}
\begin{equation}\label{Flory}
N_x = N(1-p)p^{x-1}.
\end{equation}

We now compare this distribution to the results from our numerical solution of Eqs.~(\ref{F}), (\ref{kinetic-eq}), and (\ref{Fbeta}) for $\bar{\psi}=-0.4$ (see Fig.~\ref{timeseries04beta10}). For this purpose, we consider each pixel of $\psi\geq0$ as holding a unit. To determine $N_x$, we extracted at each time the number of threads that consist of $x$ connected pixels of $\psi\geq0$. $N$ is given by the total number of threads (connected objects of $\psi\geq0$) at this time. In order to identify the total number of all units $N_0$, we counted the total number of pixels of $\psi\geq0$ at a late reaction time, which was at $t=800$. 

The distributions were determined from nine separate numerical samples of different initial fluctuations around the spatially homogeneous state. Each sample was of size $512\times512$ grid points, the parameter values were chosen as given by the caption of Fig.~\ref{timeseries04beta10}. For small values of $x$, we expect a significant deviation of $N_x$ from our numerical results. The reason is that our process of phase separation does not support objects of a diameter smaller than $d$. Also, diffusion can be important for small objects, before the process of chain connections becomes dominant. 

\begin{figure}
\includegraphics[width=8.5cm]{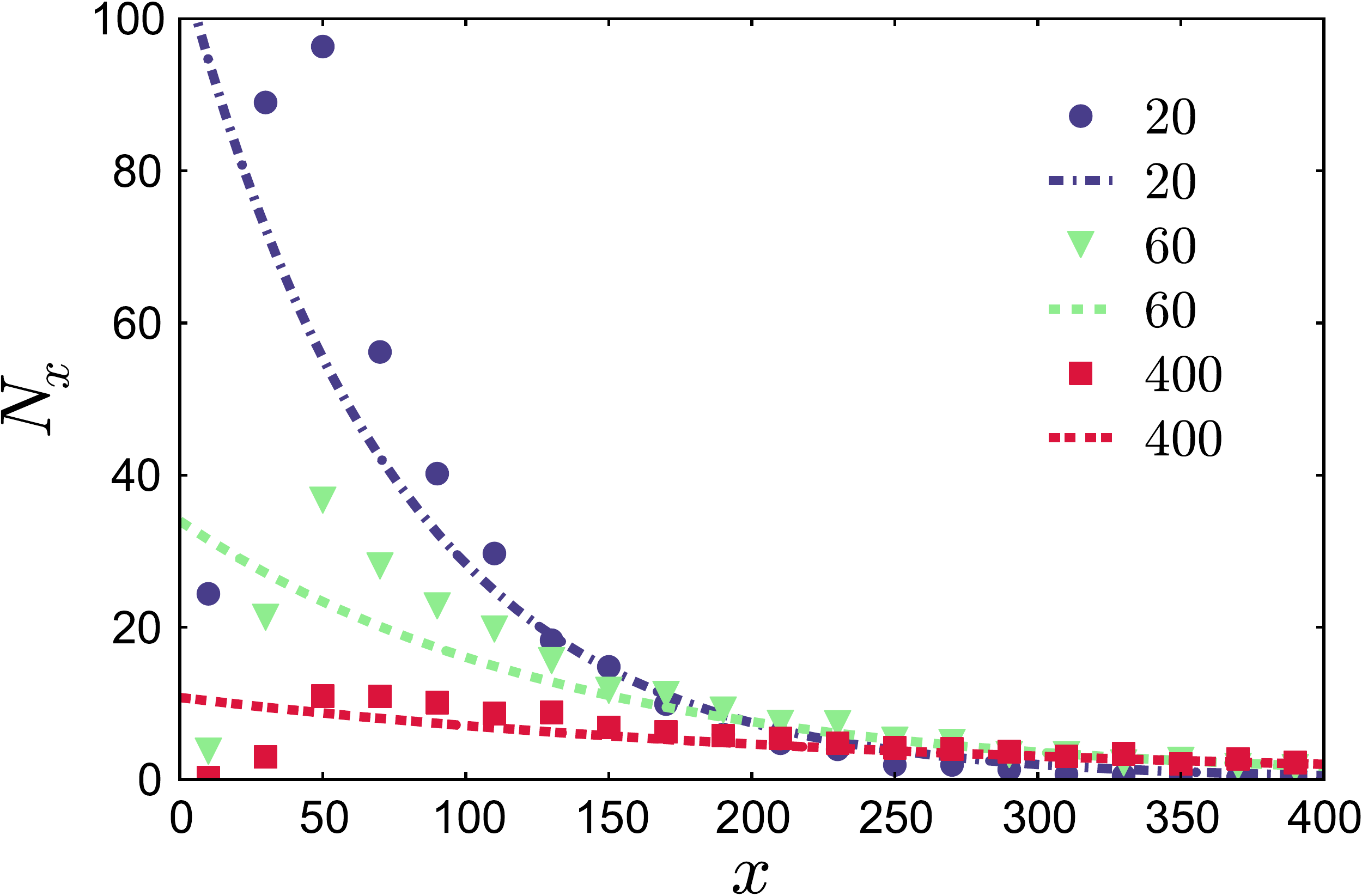}
\caption{(Color online). Time evolution of the chain/thread length distribution $N_x$, where $x$ gives the number of units in the chain/thread. The numerical data points were obtained from nine independent numerical samples at the times indicated on the upper right. Corresponding parameter values are given by the caption of Fig.~\ref{timeseries04beta10}. Each point is an average over $20$ sizes. The lines follow from the Flory chain length distribution for step-growth polymerizations as given by Eq.~(\ref{Flory}). The necessary parameters were extracted from the numerical results.}
\label{laengenvert}
\end{figure}

Fig.~\ref{laengenvert} compares the results from our numerical solutions to the predictions of Eq.~(\ref{Flory}). 
The numerical data points are averaged over $20$ object sizes. We chose this size of averaging because it corresponds to the minimal supported object size of $\pi(d/2)^2$. Clearly, we can see the deviations outlined above. Considering, however, the fact that there is no adjustable parameter, the description by Eq.~(\ref{Flory}) works reasonably well. It allows an approximate connection to step-growth polymerization processes. 

In the second part of this section, we explain that a concentration field interpretation of Eqs.~(\ref{F}), (\ref{kinetic-eq}), and (\ref{Fbeta}) can be seen as a phenomenological approach to qualitatively describe micellar, vesicular, and bilayer membrane systems. For this purpose, we think, e.g., of surfactant molecules in an aqueous solution. The surfactant molecules consist of a hydrophilic head group and a hydrophobic tail. We assume that the mass density of the hydrophilic and hydrophobic parts is identical. Then we can characterize the pure hydrophilic components (water and hydrophilic head groups) by a concentration $\psi=-1$, and the pure hydrophobic parts by $\psi=+1$. 

The phase separation for high enough values of $\vartheta$ corresponds to the separation of hydrophilic and hydrophobic components. For example, in Fig.~\ref{streifen}(a), the spheres would correspond to micelles, and the stripes to bilayer membranes. The hydrophilic head groups are on average located on the dark side ($\psi\approx-1$) of the dark-to-bright boundaries. On the contrary, the hydrophobic tails point into the center of the bright regions ($\psi\approx+1$). 

Since hydrophilic and hydrophobic parts are connected to each other, the width of the hydrophobic region can at maximum be twice the average length of the hydrophobic tails. We therefore roughly identify the thickness $d$ of the bright regions in Fig.~\ref{streifen}(a) as twice the length of the hydrophobic tails. The $\beta$-term in Eq.~(\ref{Fbeta}) determines the degree of depletion of hydrophobic groups around the regions of hydrophilic head groups. 

It may be possible to relate processes similar to the ones shown in Fig.~\ref{bildausschnitte04d15} to the formation of a vesicle. In the early stages of the numerical calculations, or throughout the calculations of section \ref{stripes}, the threads would have to be interpreted as wormlike micelles. The higher curvature energy density at the ends of wormlike micelles can be mimicked by our energy functional, as, e.g., depicted in the upper picture of Fig.~\ref{energdens}. In the following, however, we will restrict ourselves to qualitative hydrodynamic considerations. More precisely, we investigate the flow behavior of vesicular structures in simple shear and pipe flows. 

For this purpose, we coupled Eqs.~(\ref{F}), (\ref{kinetic-eq}), and (\ref{Fbeta}) with the Navier-Stokes equation, which leads to \cite{bahiana1990cell}
\begin{eqnarray}
  \frac{\partial\psi}{\partial t} & = & -\mathbf{v}\cdot\mathbf{\nabla}\psi 
    +L\Delta\frac{\delta{\cal F}}{\delta\psi} \label{NSpsi},\\
  \frac{\partial\mathbf{v}}{\partial t} & = & 
    -\mathbf{\nabla}p -\mathbf{v}\cdot\mathbf{\nabla}\mathbf{v} +\eta\Delta\mathbf{v}
    -(\psi-\bar{\psi})\mathbf{\nabla}\frac{\delta{\cal F}}{\delta\psi}. 
  \label{NSv} 
\end{eqnarray}
Here, we have assumed incompressibility. This makes the continuity equation reduce to $\nabla\cdot\mathbf{v}=0$. $L$ is the same parameter as in Eq.~(\ref{kinetic-eq}), and $\eta$ denotes the viscosity. The (constant) mass density $\rho$ has been scaled out. 

We set the parameter values to $\vartheta=5$, $\alpha=3$, $d=15.3$, $\beta=0.3$, $b=10.2$, and $\eta=100$. In the case of simple shear flow, we set $L=10$, whereas we chose $L=20$ in the case of pipe flow. The numerical calculation grid was of size $256\times128$ sites with a lattice distance of $dx=1$, and the size of the time steps was $dt=0.00005$. We chose periodic boundary conditions at the left and right edges of our rectangular calculation grid. On the contrary, noflux boundary conditions were imposed at the top and bottom boundaries. The flow direction will thus be horizontal. 

To enforce steady simple shear flow, we set the horizontal velocity equal to $-2$ at the upper boundary. In all other cases, both velocity components were set equal to zero at the top and bottom boundaries. To induce steady horizontal pipe flow, we added a constant gravity-like volume force $\mathbf{g}=(0,g)$ to Eq.~(\ref{NSv}), where $g<0$ in our case. The horizontal flow will therefore be directed towards the left edge of the calculation grid in both geometries. We used a MAC scheme on a staggered grid to numerically integrate Eqs.~(\ref{NSpsi}) and (\ref{NSv}) forward in time \cite{fletcher1991computational}. The intermediate Laplace equation resulting for the pressure from the incompressibility condition $\nabla\cdot\mathbf{v}=0$ was solved through a spectral method \cite{press1992numerical}. Third and first order upwind schemes were used to discretize the advective term in Eqs.~(\ref{NSpsi}) and (\ref{NSv}), respectively. 

In order to start our investigations, we provided an annulus of $\psi=+1$ and thickness $d$ in an environment of $\psi=-1$. We let it relax to a spherical vesicle-like object by numerically iterating Eq.~(\ref{kinetic-eq}) for a time interval $\Delta t=40$. These vesicles were then used as an initial condition for the numerical investigations of Eqs.~(\ref{NSpsi}) and (\ref{NSv}). The objects that we describe by this sort of concentration field approach have to be classified as fluid vesicles. That is, viscous flow of the constituents is possible within the vesicle membranes. No elastic shear forces occur within the local membrane planes, which would be the case for elastic membranes. 

\begin{figure}
\includegraphics[width=8.5cm]{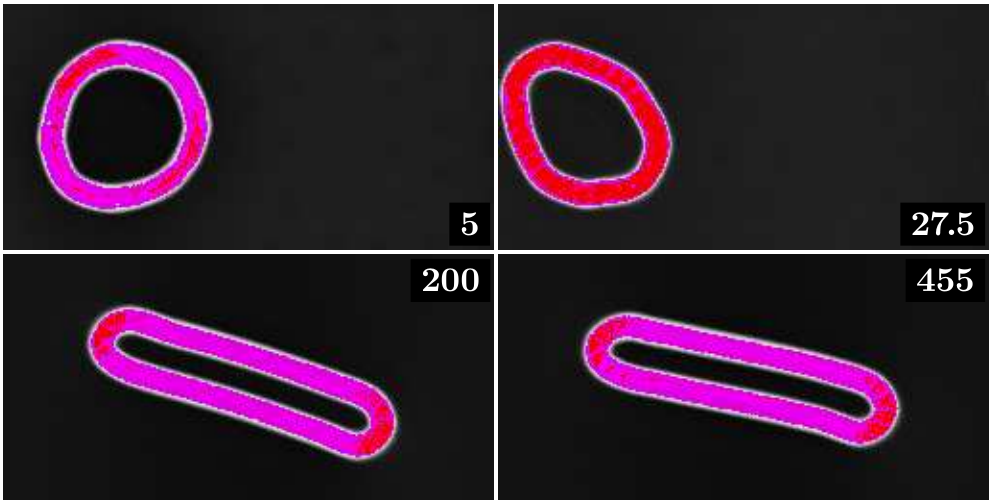}
\caption{(Color online). Simple shear flow of a vesicle characterized by Eqs.~(\ref{NSpsi}) and (\ref{NSv}). An approximately spherical vesicle was provided as an initial condition. At time $t=5$ the shear was started by setting the horizontal velocity component at the upper boundary to a nonzero value, pointing to the left, the effect of which is clearly visible at time $t=27.5$. The bottom row shows the vesicle after having crossed the lateral periodic boundaries once (time $t=200$) and twice (time $t=455$). Pink color: $\psi>0.8$, red color: $\psi>1$. Parameter values were set to $\vartheta=5$, $\alpha=3$, $d=15.3$, $\beta=0.3$, $b=10.2$, $\eta=100$, and $L=10$. The grid size was $256\times128$ sites with a lattice distance of $dx=1$, and the size of the time steps was $dt=0.00005$.}
\label{schervesikel}
\end{figure}

We give an example for the obtained behavior of a vesicle under shear flow in Fig.~\ref{schervesikel}. 
The first picture shows the vesicle at the time when the shear flow is started by setting the horizontal velocity component to a nonzero value at the upper boundary. In the second picture, the effect of this shear deformation starts to become visible. The bottom row of Fig.~\ref{schervesikel} depicts the vesicle as it is strongly deformed by the flow field of steady shear. We show its state after having crossed the periodic left-right boundary once and twice. 

The long axis of the stretched vesicle approaches a steady orientation with respect to the flow direction. This behavior is well known from numerical studies, e.g.~solvent multiparticle collision dynamics combined with coarse-grained membrane models \cite{noguchi2004fluid,messlinger2009dynamical}, as well as experimental observations \cite{abkarian2002tank,kantsler2005orientation}. It is referred to as the tank-treading regime, because it is usually accompanied by a rotational motion of the vesicle membrane around its interior. 

We must remark in this context that our characterization in its current implementation does not reproduce an essential feature of the experimental systems. Real vesicles are practically impermeable for the surrounding solvent on experimental time scales. Generally, in field descriptions, this can be achieved by setting the membrane velocity equal to the local fluid flow velocity \cite{beaucourt2004steady}. In our case, this corresponds to letting $L\rightarrow0$ in Eq.~(\ref{NSpsi}). Then the concentration field describing the location of the membrane is advected by the flow velocity only and does not diffuse. However, this limit is not accessible by our current numerical scheme and is therefore beyond the scope of this work. 

\begin{figure}
\includegraphics[width=8.5cm]{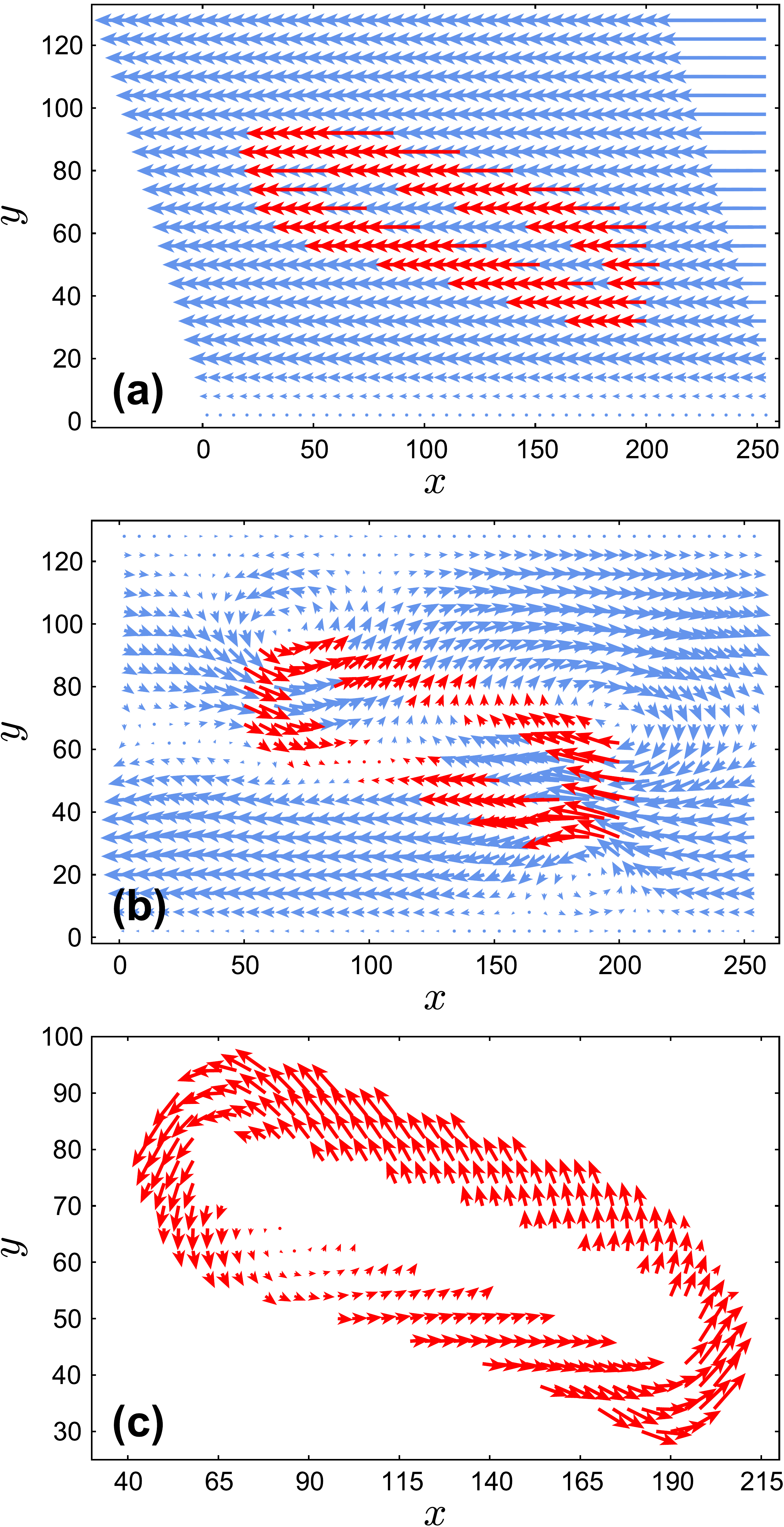}
\caption{(Color online). Flow field $\mathbf{v}$ corresponding to the case of simple shear in Fig.~\ref{schervesikel} at time $t=455$. The red color indicates the flow field within the membrane ($\psi>0.5$), the blue color the flow field within the remaining solvent. (a) Steady shear flow acting on the membrane. (b) Subtraction of the linear shear profile reveals the fine structure and distortion of the flow field due to the presence of the membrane. Four vortices form, two at the front and two at the back of the membrane (the bottom one at the front is not as clearly visible as the other ones). Competition within the group of the two front vortices and within the group of the two vortices at the back is reflected by the finite inclination angle of the membrane. (c) Tendency of tank-treading in the rest frame of the membrane. For clarity, the horizontal velocity component is rescaled by a factor $0.01$ w.r.t.~the vertical one.}
\label{tanktread}
\end{figure}

Analyzing the flow field around the vesicle, we do, however, observe the tendency of tank-treading along the vesicle membrane. Fig.~\ref{tanktread}(a) shows the velocity profile when the membrane is exposed to a steady shear flow. The situation corresponds to the last picture of Fig.~\ref{schervesikel}. 
We find the fine structure of the flow velocity field by subtracting the linear shear profile, as depicted in Fig.~\ref{tanktread}(b). This reveals how the membrane influences the velocity profile of the surrounding fluid. Two vortices form at the front (left) and two at the back (right) of the vesicle (the bottom one at the front is not as clearly visible as the others). Within these two groups of vortices, one vortex adds to aligning the elongated vesicle along the flow direction, the other to aligning it perpendicular to the flow field. This competition is reflected by the finite inclination angle of the stretched vesicle. Furthermore, the membrane connects the flow in the upper half of the channel to the flow in the lower half and slightly evens the different flow velocities. In Fig.~\ref{tanktread}(b), this becomes clear from the velocity field predominantly pointing to the right in the upper part, whereas it is oriented toward the left in the lower part. Generally, such a balancing effect between the areas of different flow velocities can be connected to the tank-treading motion of the membrane. Fig.~\ref{tanktread}(c) displays the flow field of the membrane in its rest frame. A rotational tank-treading-like motion of the membrane is revealed by rescaling the horizontal component of velocity by a factor of $0.01$ with respect to the vertical component. 

Apart from the tank-treading motion, at least two other regimes were identified numerically and experimentally \cite{noguchi2004fluid,messlinger2009dynamical, abkarian2002tank,kantsler2005orientation,kantsler2006transition}. If the inclination angle of the long axis of the elongated vesicle is not steady but varies in time, the motion is called tumbling. In addition, a more irregular dynamic mode of motion was identified: it includes tumbling, features dynamic shape deformations away from the purely elongated state, and was named swinging mode. 

Mainly two parameters were found to determine the mode of motion. First, this is the reduced volume. It measures the deviation from a spherical shape by relating the surface to the volume of the vesicle. The second parameter is the ratio between the fluid viscosities inside and outside the vesicle. Imposing impermeability of the membrane is crucial to control these parameters. It will therefore be an important step for future work on our description. 

\begin{figure}
\includegraphics[width=8.5cm]{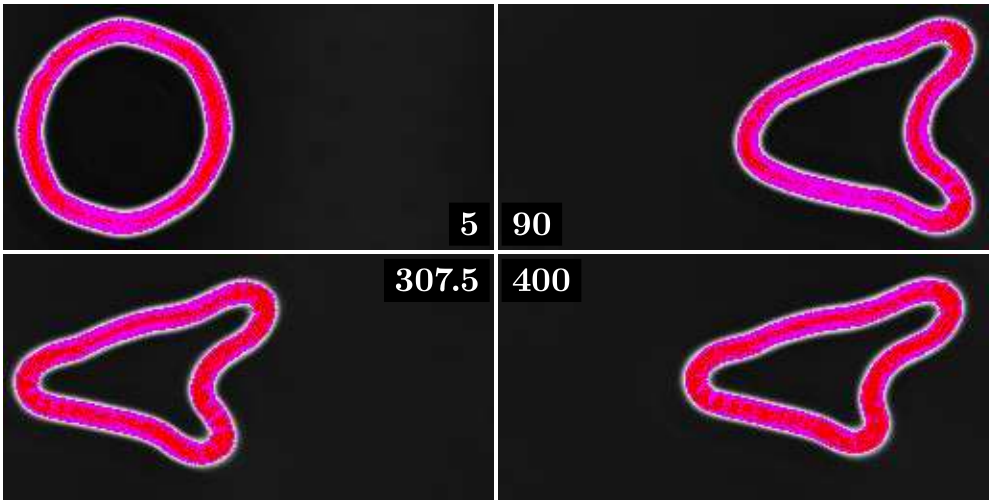}
\caption{(Color online). Time series of a vesicle exposed to pipe flow. First, the vesicle folds into a parachute shape, following the Poiseuille flow profile. Then it tends to elongate along the flow direction. Parameter values and color code as given by the caption of Fig.~\ref{schervesikel}, except for $L=20$ and $g=-0.4$.}
\label{parachute}
\end{figure}

We close this section by pointing out our qualitative results for the pipe flow geometry. Fig.~\ref{parachute} shows an initially spherical vesicle that has a size of the order of the channel diameter. 
When the flow starts, the vesicle is deformed according to the profile of the Poiseuille-like flow. It folds into a parachute-like shape. At longer times, we find a transition to a shape that is elongated in the flow direction (compare Ref.~\cite{noguchi2005shape}). 

\begin{figure}
\includegraphics[width=8.5cm]{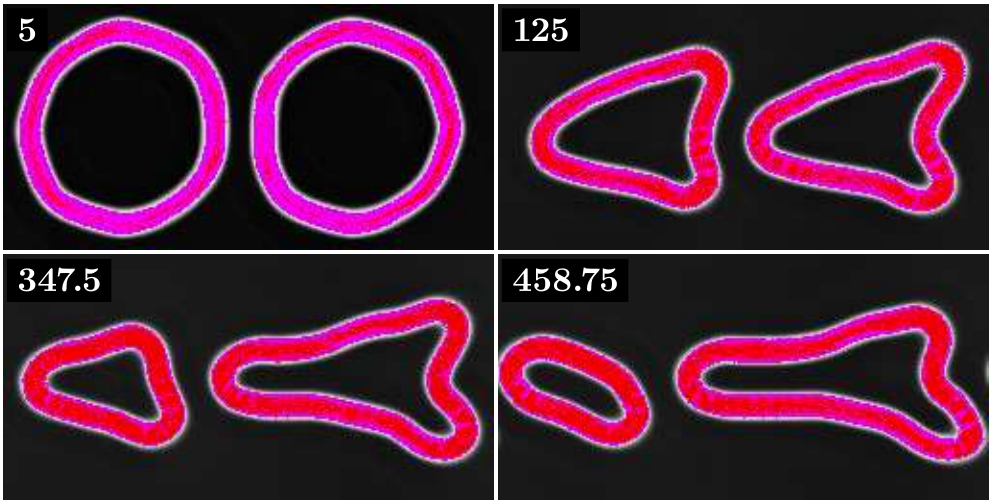}
\caption{(Color online). Two vesicles in a pipe geometry. As time proceeds, the rear vesicle grows on cost of the front vesicle. Parameter values and color code as given by the caption of Fig.~\ref{schervesikel}, except for $L=20$ and $g=-0.5$.}
\label{doublepipe}
\end{figure}

Finally, we put two vesicles into the channel, as depicted by Fig.~\ref{doublepipe}. 
Initially, they show the same behavior as the single vesicle in Fig.~\ref{parachute}. However, as time proceeds, the front vesicle looses material that is collected by the rear one. Consequently, the first vesicle shrinks, whereas the second one grows. It would be interesting to know whether a corresponding process can be observed in an experimental system.


\section{\label{conclusion}Conclusions}

We have introduced a simple nonlocal energy density in Eq.~(\ref{F}) that was discretized for numerical investigations as explained in the appendix. Starting from the disordered (spatially homogeneous) state, this potential describes the formation of non-periodic localized thread-like structures. These structures are metastable, and finally we observe a macroscopic phase separation. The scaling behavior of the latter coarse-graining process is similar to the case of the Cahn-Hilliard model. We could omit the macroscopic phase separation by adding expression (\ref{Fbeta}) to the energy density. It stabilizes the localized, non-periodic thread-like stripe structures. In this case, we observe a connecting process of different threads to reduce the number of high-energetic end points of the threads. Our analysis revealed a regime during which the number of end points of the threads decays exponentially. Finally, we discussed two possible applications of our model in the field of soft matter. On the one hand, this was a density field description of polymers, where for the moment we concentrated on the chain length distribution function during the formation process. On the other hand, we focussed on the hydrodynamic properties of vesicular structures. Simple shear and pipe flow geometries were qualitatively addressed, where our description reproduces generic features of the flow behavior. 

As pointed out, several aspects remain for future investigations. We have indicated that the variation of the various parameter values involved in the description leads to further qualitatively different textures. A complete phase diagram needs to be established. Next, to allow more quantitative studies, our numerical discretization scheme should be improved. For example, a further decoupling from the underlying grid directions might be achieved through an implementation on a hexagonal lattice. Furthermore, a generalization to three spatial dimensions would be desirable. Lastly, we have followed a phenomenological approach so far. It would be interesting to investigate whether our functional can be related to more microscopic approaches through appropriate coarse-graining, and which changes to the functional form this would imply.

\begin{acknowledgments}
The author thanks Nigel Goldenfeld and Takao Ohta for stimulating discussions on related fields, and Nigel Goldenfeld for a stay in his group, where this work was performed. This work was supported by the Deutsche Forschungsgemeinschaft through a research fellowship, which is gratefully acknowledged. 
\end{acknowledgments}

\appendix*

\section{Discretized numerical evaluation of the functional derivative and kinetic equation}

In this appendix we describe our numerical evaluation of the functional derivative in Eq.~(\ref{kinetic-eq}). The form of the corresponding energy density is given by Eq.~(\ref{F}). 

Due to the nonlocal contribution in Eq.~(\ref{F}), it is not straightforward to perform the functional derivative analytically. We use the following definition of the functional derivative:
\begin{equation}\label{varablcont}
\frac{\delta\mathcal{F}\!\left\{\psi(\mathbf{r'})\right\}}{\delta\psi(\mathbf{r})} = 
\lim_{\epsilon\rightarrow0}\frac{1}{\epsilon}\Big[ 
\mathcal{F}\left\{\psi(\mathbf{r'})+\epsilon\,\delta(\mathbf{r'}-\mathbf{r})\right\}-
\mathcal{F}\left\{\psi(\mathbf{r'})\right\} \Big].
\end{equation}
This is the correct starting point for the transformation from the analytic continuum picture to the discrete numerical calculation. 

The numerical lattice is of finite size and defines a discrete system. We use a rectangular domain on a square lattice of $N_x\times N_y$ lattice sites with periodic boundary conditions. The lattice sites are indexed by pairs $(i,j)$, where $i=0,\dots,N_x-1$ and $j=0,\dots,N_y-1$. Consequently, the continuous functional form $\mathcal{F}\left\{\psi(\mathbf{r})\right\}$ transforms to a simple multivariate function $\mathcal{F}\left[\psi_{ij}\right]$ of $N_x\times N_y$ variables $\psi_{ij}$.

In analogy, the Dirac delta function $\delta(\mathbf{r'}-\mathbf{r})$ is now given by a product of Kronecker deltas $\delta_{il}\delta_{jm}$, where $(i,j)$ and $(l,m)$ correspond to the positions $\mathbf{r}$ and $\mathbf{r'}$, respectively. Therefore, the functional derivative in Eq.~(\ref{varablcont}) transforms to a simple partial derivative,
\begin{equation}\label{varabldis}
\frac{\partial\mathcal{F}\!\left[\psi_{lm}\right]}{\partial\psi_{ij}} = 
\lim_{\epsilon\rightarrow0}\frac{1}{\epsilon}\Big[ 
\mathcal{F}\left[\psi_{lm}+\epsilon\,\delta_{il}\delta_{jm}\right]-
\mathcal{F}\left[\psi_{lm}\right] \Big].
\end{equation}
This is the expression that we use to numerically evaluate the functional derivatives. 

Furthermore, we discretize the gradient field in Eq.~(\ref{F}) as
\begin{equation}\label{gradientdis}
\nabla\psi\,|_{ij}= \left( \frac{\psi_{i+1,j}-\psi_{ij}}{dx}, 
\frac{\psi_{i,j+1}-\psi_{ij}}{dx} \right).
\end{equation}
We can motivate this choice by considering a quadratic gradient contribution $\frac{1}{2}(\nabla\psi\,|_{\mathbf{r}})^2$ in the energy density, as it occurs, e.g., in the energy functional for the Cahn-Hilliard model \cite{cahn1958free}.
On the one hand, analytically, the functional derivative of this term follows immediately as $-\Delta\psi\,|_{\mathbf{r}}$. On the other hand, the discretization of the term $\frac{1}{2}(\nabla\psi\,|_{\mathbf{r}})^2$ according to Eq.~(\ref{gradientdis}) reads $\frac{1}{2}\left[(\psi_{i+1,j}-\psi_{ij})^2+(\psi_{i,j+1}-\psi_{ij})^2 \right]/dx^2$. We can perform the functional derivative as described by Eq.~(\ref{varabldis}). Equating both results at position $|_{\mathbf{r}}\equiv|_{ij}$ yields
\begin{equation}
\Delta\psi\,|_{ij} = \frac{\psi_{i+1,j}+\psi_{i-1,j}+\psi_{i,j+1}+\psi_{i,j-1}
-4\psi_{ij} }{ dx^2}.
\end{equation}
As we can see, we correctly obtain the discretization of the Laplace operator as given by the stencil
\begin{equation}
\begin{array}{|c|c|c|}
              \hline
0 &  1 & 0 \\ \hline
1 & -4 & 1 \\ \hline
0 &  1 & 0 \\ \hline
\end{array}.
\end{equation}
This discretization of the Laplacian introduces anisotropy \cite{patra2006stencils}. More sophisticated discretizations than expression (\ref{gradientdis}) must be worked out for the gradient field when a fully isotropic discretization of the Laplacian operator should be obtained from the functional derivative. This is beyond the scope of the considerations in this paper. 

In order to write our results in a compact way, we define an abbreviation for the discretized nonlocal subscripts $|_{\left. \mathbf{r}+d\frac{\nabla\psi}{\|\nabla\psi\|}\right|_{ij}}$ at site $(i,j)$:
\begin{equation}\label{pointerdef}
|_{xp_{ij},yp_{ij}}\equiv |_{ \left. \mathbf{\hat{x}}\cdot\left( {\mathbf{r}+d\frac{\nabla\psi}{\|\nabla\psi\|}} \right) \right|_{ij},\, \left. 
\mathbf{\hat{y}}\cdot\left( {\mathbf{r}+d\frac{\nabla\psi}{\|\nabla\psi\|}} \right) \right|_{ij} }.
\end{equation}
The pair $(xp_{ij},yp_{ij})$ indexes the lattice site that the subscript $|_{\mathbf{r}+d\frac{\nabla\psi}{\|\nabla\psi\|}}$ points to (the ``$p$'' stands for ``pointer''). Here, the gradients in the subscript are discretized according to Eq.~(\ref{gradientdis}).

On the one hand, the value $\psi_{ij}$ contributes to the value of the energy density at the grid site $(i,j)$. This comes from the local $\vartheta$-term and from the local term $\nabla\psi|_{\mathbf{r}}$ in the square brackets of Eq.~(\ref{F}). It is crucial to note that, on the other hand, $\psi_{ij}$ contributes to the value of the energy density at all other grid sites $(l,m)$ that point to $(i,j)$ through the subscript $|_{xp_{lm},yp_{lm}}\equiv|_{\mathbf{r'}+d\frac{\nabla\psi}{\|\nabla\psi\|}}$. The latter part comes from the nonlocal term $\nabla\psi|_{\mathbf{r}+d\frac{\nabla\psi}{\|\nabla\psi\|}}$ in the square brackets of Eq.~(\ref{F}). 

Consequently, at each time step, we must determine for each lattice site $(i,j)$ all other lattice sites $(l,m)$ that point to $(i,j)$; i.e.~all other lattice sites $(l,m)$ satisfying the condition $(xp_{lm},yp_{lm})=(i,j)$. In the supplemental material \cite{supplemental} we present a scheme to perform this task at an overall computational cost of linear order, $\mathcal{O}(N_xN_y)$. We used dynamic list structures for this purpose. 

Following the procedure outlined in Eq.~(\ref{varabldis}), we obtain the contributions to the functional derivative at site $(i,j)$. From the $\vartheta$-term in Eq.~(\ref{F}), we directly find 
\begin{equation}
  \vartheta\left(\psi_{ij}^3-\psi_{ij}\right).
\end{equation} 

Next, at site $(i,j)$, the local term $\nabla\psi|_{\mathbf{r}}$ leads to a contribution
\begin{eqnarray}
\frac{2\alpha}{dx^2}\sum_{l,m} 
  & \Big\{ &
  (\psi_{l+1,m}-\psi_{lm}+\psi_{xp_{lm}+1,yp_{lm}}-\psi_{xp_{lm},yp_{lm}}) 
  \nonumber\\
&& \times (\delta_{l+1,i}\delta_{mj}-\delta_{li}\delta_{mj}) \nonumber\\[.4cm]
&+& 
  (\psi_{l,m+1}-\psi_{lm}+\psi_{xp_{lm},yp_{lm}+1}-\psi_{xp_{lm},yp_{lm}})
  \nonumber\\[.2cm]
&& \times (\delta_{li}\delta_{m+1,j}-\delta_{li}\delta_{mj}) \;\;\Big\}.  
\end{eqnarray}

For the nonlocal term $\nabla\psi|_{\mathbf{r}+d\frac{\nabla\psi}{\|\nabla\psi\|}}$, we obtain a contribution
\begin{eqnarray}
\frac{2\alpha}{dx^2}\sum_{l,m} 
  & \Big\{ &
  (\psi_{l+1,m}-\psi_{lm}+\psi_{xp_{lm}+1,yp_{lm}}-\psi_{xp_{lm},yp_{lm}}) 
  \nonumber\\
&& \times 
    (\delta_{xp_{lm}+1,i}\delta_{yp_{lm},j}-\delta_{xp_{lm},i}\delta_{yp_{lm},j}) 
  \nonumber\\[.4cm]
&+& 
  (\psi_{l,m+1}-\psi_{lm}+\psi_{xp_{lm},yp_{lm}+1}-\psi_{xp_{lm},yp_{lm}})
  \nonumber\\[.2cm]
&& \times 
  (\delta_{xp_{lm},i}\delta_{yp_{lm}+1,j}-\delta_{xp_{lm},i}\delta_{yp_{lm},j}) 
  \;\; \Big\}  
\end{eqnarray}
to the functional derivative at site $(i,j)$. 

Finally, we have to add the contribution arising from the term with the coefficient $\beta$ in Eq.~(\ref{Fbeta}). It turns out that the gradients in the subscript $|_{\mathbf{r}-b\frac{\nabla\psi}{\|\nabla\psi\|}}$ must be discretized differently than in Eqs.~(\ref{gradientdis}) and (\ref{pointerdef}). Otherwise the discrete nature of the numerical calculation leads to a systematic artificial drift. 

In analogy to Eq.~(\ref{pointerdef}), we define
\begin{equation}\label{pointerbetadef}
|_{xp\beta_{ij},yp\beta_{ij}}\equiv |_{ \left. \mathbf{\hat{x}}\cdot\left( {\mathbf{r}-b\frac{\nabla\psi}{\|\nabla\psi\|}} \right) \right|_{ij},\, \left. 
\mathbf{\hat{y}}\cdot\left( {\mathbf{r}-b\frac{\nabla\psi}{\|\nabla\psi\|}} \right) \right|_{ij} }.
\end{equation}
Here, however, the gradients in the subscript are discretized according to 
\begin{equation}
\nabla\psi\,|_{ij}= \left( \frac{\psi_{i+1,j}-\psi_{i-1,j}}{2dx}, 
\frac{\psi_{i,j+1}-\psi_{i,j-1}}{2dx} \right)
\end{equation}
instead of Eq.~(\ref{gradientdis}). For each site $(l,m)$ that satisfies the condition $(xp\beta_{lm},yp\beta_{lm})=(i,j)$ we obtain a contribution
\begin{equation}
2\beta(\psi_{ij}+1)
\end{equation}
to the functional derivative at site $(i,j)$. 

We present a schematic implementation of the above results using a C-like pseudocode in the associated supplemental material \cite{supplemental}.

\end{document}